\documentstyle[preprint,aps,epsf]{revtex}
\newcommand{\be}{\begin{equation}}
\newcommand{\ba}{\begin{eqnarray}}
\newcommand{\ee}{\end{equation}}
\newcommand{\ea}{\end{eqnarray}}
\newcommand{\nn}{\nonumber}

\newcommand{\ovl}{\overline}

\newcommand{\sla}[1]{#1 \!\!\! \slash}
\renewcommand{\theequation}{\arabic{section}.\arabic{equation}}
\def\E{e^+e^-}
\def\Emm{e^-e^-}
\def\Epp{e^+e^+}
\def\Eetc{e^{\pm}e^-W^{\mp}W^+}
\begin{document}
\draft
\preprint{PITHA 96/5}
\title{Pair Production and Correlated Decay\\
 of Heavy Majorana Neutrinos in $e^+e^-$ Collisions}

\author{Axel Hoefer\footnote{email: hoefer@physik.rwth-aachen.de} and 
L.~M.~Sehgal\footnote{email: sehgal@physik.rwth-aachen.de}}
\address{Institut f\"ur Theoretische Physik (E), RWTH Aachen\\
D-52074 Aachen, Germany}

\maketitle
\begin{abstract}
We consider the process $e^+e^-\to N_1N_2$, where $N_1$ and $N_2$
are heavy Majorana particles, with relative $CP$ given by $\eta_{CP}=+1$
or $-1$, decaying subsequently via $N_1,N_2\to W^{\pm}e^{\mp}$.
We derive the energy and angle correlation of the dilepton final state,
both for like-sign ($e^{\mp}e^{\mp}$) and  unlike-sign ($e^-e^+$)
configurations. Interesting differences are found between the cases
$\eta_{CP}=+1$ and $-1$. The characteristics of unlike-sign $e^+e^-$
dileptons originating from a Majorana pair $N_1N_2$ are contrasted
with those arising from the reaction $e^+e^-\to N\bar{N}\to W^+e^-W^-e^+$, 
where $N\bar{N}$ is a Dirac particle-antiparticle pair.
\end{abstract}
\newpage
\section{introduction}
In an interesting paper \cite{Tsai}, Kogo and Tsai have analysed the
reaction $e^+e^-\to N_1N_2$, where $N_{1,2}$ are heavy Majorana neutrinos,
and compared the cases where the relative $CP$ of $N_1$ and $N_2$ is
$\eta_{CP}=+1$ and $-1$.
It was found that the two cases differ in threshold behaviour, in
angular distribution, and in the dependence on the spin-directions of
$N_1$ and $N_2$. A comparison was also made between the Majorana process
and the Dirac process $e^+e^-\to N\bar{N}$, where $N\bar{N}$ is a 
Dirac particle-antiparticle pair. A related analysis was carried out in 
Ref.~\cite{Ma}. (The contrast between Majorana neutrinos
and Dirac neutrinos has been the subject of several other papers
(e.g.~\cite{Maal,Lan,Denner,Buch}) and monographs (\cite{Moha,Kayser1}))

In the present paper, we examine how the differences between the cases
$\eta_{CP}=+1$ and $-1$ propagate to the decay products of $N_1$ and $N_2$,
assuming the decays to take place via $N_{1,2}\to W^{\pm}e^{\mp}$.
We focus on the like-sign lepton pair created in the reaction chain
$e^+e^-\to N_1N_2\to W^+e^-W^+e^-$, which
is a characteristic signature of Majorana pair production.
We derive, in particular, the correlation in the energies of the
$e^-e^-$ pair, and in their angles relative to the $e^+e^-$ axis.
Interesting differences are found between the cases $\eta_{CP}=+1$ and $-1$.
We also examine the behaviour of the unlike-sign dileptons $e^+e^-$,
comparing the Majorana cases with dileptons created in the production and 
decay of a Dirac $N\bar{N}$ pair, i.e. $e^+e^-\to N\bar{N}\to W^+e^-W^-e^+$.

\section{Characteristics of the Reaction $\E \to N_1N_2$}
The analysis of Ref.~\cite{Tsai} was carried out in the context of 
the simple production mechanism for $\E\to N_1N_2$ shown in Fig.~1,
and we begin by recapitulating the essential results. 
The interaction Lagrangian is taken to be
\ba
{\cal L}_1(x)\; = \;-\; \frac{g}{2\; cos\;\theta_W}&\biggl [ &
      \bar{e}\;(x)\;\gamma_{\mu}\;(c_V - c_A\gamma_5)\;e\;(x) \nn\\
& + &\alpha_N\;
\bar{N_1}(x)\;\gamma_{\mu}\;\frac{1}{2}(1-\gamma_5)\;N_2(x) \nn\\
& + & \alpha_N\;\bar{N_2}(x)\;\gamma_{\mu}\;\frac{1}{2}(1-\gamma_5)\;N_1(x)
\;\;\biggr ]\;\;Z^{\mu}(x)\;\;,
\ea 
where $c_V$, $c_A$ and $\alpha_N$ may be regarded as real phenomenological 
parameters. (For the standard Z-boson, $c_V = -1/2+2 \sin^2\theta_W$, 
$c_A = -1/2$). The matrix element for Majorana neutrinos 
(with momenta and spins as indicated in Fig.~1) is
\ba
{\cal M}_m\;=\;-i\alpha_N\;(\frac{g}{2\;cos\;\theta_W})^2\;\;j_{\mu}^e\;
                \;\Delta_Z^{\mu\nu}
               &\biggl[&\bar{u}_{t_1}(q_1)\;\gamma_{\nu}\;
                \frac{1}{2}(1-\gamma_5)
               \;v_{t_2}(q_2)\;\lambda_2\nn\\
          & - & \bar{u}_{t_2}(q_2)\;\gamma_{\nu}\;\frac{1}{2}(1-\gamma_5)
                  \;v_{t_1}(q_1)\;\lambda_1\;\biggr ]\;,
\ea
where
\be
j_{\mu}^e\;=\;\bar{v}_{s_2}(p_2)\;\gamma_{\mu}\;(c_V-c_A\gamma_5)
                 \;u_{s_1}(p_1)
\ee
and
\be
\Delta_Z^{\mu\nu}\;=\;\frac{g^{\mu\nu}\;-\;q^{\mu}q^{\nu}/m_Z^2}{q^2\;-
\;m_Z^2\;+i\;m_Z\;\Gamma_Z}\;\;.
\ee
Assuming $CP$-invariance the factors $\lambda_1$, $\lambda_2$ 
in Eq.~(2.2) are such that
$\lambda_1\lambda_2^*=+1(-1)$ when $N_1$ and $N_2$ have the same (opposite)
$CP$-parity \cite{Kayser2}. Rewriting the second term in Eq.~(2.2) as
\be
\bar{u}_{t_2}(q_2)\;\gamma_{\nu}\;\frac{1}{2}(1-\gamma_5)\;v_{t_1}(q_1)\;=\;
\bar{u}_{t_1}(q_1)\;\gamma_{\nu}\;\frac{1}{2}(1+\gamma_5)\;v_{t_2}(q_2)\;\;,
\ee
we observe that the current of the Majorana neutrinos is pure axial vector 
when $N_1$ and $N_2$ have the same $CP$-parity
($\eta_{CP}=\lambda_1\lambda_2^*=+1$), and pure vector when they have opposite
$CP$ ($\eta_{CP}=\lambda_1\lambda_2^*=-1$). In comparison, 
the matrix element for the Dirac process $\E\to N\bar{N}$ is
\be
{\cal M}_d\;=\;-i\alpha_N\;(\frac{g}{2\;cos\;\theta_W})^2\;j_{\mu}^e\;
\Delta_Z^{\mu\nu}
                  \;\bar{u}_{t_1}(q_1)\;\gamma_{\nu}\;\frac{1}{2}(1-\gamma_5)
                  \;v_{t_2}(q_2) \;\;.
\ee
The differential cross section for $\E\to N_1N_2$, for general masses
$m_1$ and $m_2$, and for arbitrary polarizations $\vec{n}$ and $\vec{n}'$
of the two neutrinos is given in the Appendix. In Sec.~5 we compare our
formulas with those of Ref.~\cite{Tsai}, and with special cases treated
in other papers. Here we specialise to the case 
$m_1=m_2=m_N$, for which the cross-section ($d\sigma/d\Omega$)
in the cases $\eta_{CP}=+1$ and $-1$ is
\ba
(\frac{d\sigma}{d\Omega})_+&=&\frac{1}{2}\sigma_0\;\beta^3\; 
\biggl\{\;f_1\;\bigl[(n_yn_y'-n_xn_x')S^2+(1+n_zn_z')(1+C^2)\bigr]
\;-\;f_2\;\;2(n_z+n_z')C\;\biggr\}\;\;,\\\nn\\\nn
(\frac{d\sigma}{d\Omega})_-&=&\sigma_0\;\beta\; 
\biggl\{\;f_1\;\bigl[2-\beta^2+C^2\beta^2+n_zn_z'(\beta^2+C^2(1/\gamma^2+1))
+n_xn_x'S^2(1/\gamma^2+1)\\
&&-n_yn_y'S^2\beta^2
+(n_xn_z'+n_x'n_z)2SC/\gamma^2\bigr]
\;-\;f_2\;\bigl[2(n_x+n_x')S/\gamma^2+2(n_z+n_z')C\bigr]\;\biggr\}\;.
\ea
For comparison, the differential cross section of the Dirac process
$\E\to N\bar{N}$ is
\ba
(\frac{d\sigma}{d\Omega})_d&=&\frac{1}{2}\sigma_0\;\beta\; 
\biggl\{\;f_1\;\bigl[(1+C^2\beta^2)-(n_z+n_z')\beta(1+C^2)\\\nn
&&-(n_x+n_x')SC\beta/\gamma+n_zn_z'(C^2+\beta^2) 
+(n_xn_z'+n_zn_x')SC/\gamma+n_xn_x'S^2/\gamma^2\bigr]\;+\;\\\nn 
&&f_2\;\bigl[2C\beta-(n_z+n_z')C(1+\beta^2)
-(n_x+n_x')S/\gamma+2n_zn_z'C\beta+(n_xn_z'+n_x'n_z)S\beta/\gamma\bigr]\;
\biggr\}.\nn
\ea
The symbols in Eqs.~(2.7)--(2.9) are defined as follows:
\ba
\sigma_0 &=& \frac{G_F^2\alpha_N^2}{512\pi^2}\;
\left|\frac{m_Z^2}{s-m_Z^2+im_z\Gamma_Z}\right|^2\;s\quad,\quad
\beta = (1-4m_N^2/s)^{1/2}\quad,\quad\gamma=(1-\beta^2)^{-1/2}\nn\\\\
C & = & \cos\theta\quad,\quad S=\sin\theta\quad,\quad
f_1  =  2\;(c_V^2+c_A^2)\quad,\quad f_2=4\;c_Vc_A\quad,\nn
\ea
$\theta$ being the scattering angle of $N_1$ (or $N$) with respect to
the initial $e^-$ direction. The co-ordinate axes are defined so that
the momentum- and spin-vectors of $N_1$ and $N_2$ in the $\E$ c.m.
frame have the components
\ba
q_1^\mu &=& (\gamma m,0,0,\gamma\beta m)\;\;,\;\;
t_1^\mu  =   (\gamma\beta n_z,n_x,n_y,\gamma n_z)
\;\;,\\
q_2^\mu &=& (\gamma m,0,0,-\gamma\beta m)\;\;,\;\;
t_2^\mu  =  (-\gamma\beta n_z',n_x',n_y',\gamma n_z')
\;\;.\nn
\ea   
Inspection of Eqs.~(2.7)--(2.9) reveals several interesting features:
\begin{description}
\item[(a)]  The Majorana cases '$+$' and '$-$' have different
dependence on the spin-vectors $\vec{n}$ and $\vec{n}'$, and different
angular distributions, even after the spins $\vec{n}$ and $\vec{n}'$
are summed over. These differences stem from the fact that the 
matrix element ${\cal M}_m$ in Eq.~(2.2) effectively involves an axial
vector current $\bar{N_1}\gamma_{\mu}\gamma_5 N_2$ when 
$\lambda_1\lambda_2^*=+1$ and a vector current 
$\bar{N_1}\gamma_{\mu}N_2$ when $\lambda_1\lambda_2^*=-1$.
\item[(b)]  The Majorana cases '$+$' and '$-$' differ from the Dirac case
'd', in which the current of the $N\bar{N}$-pair has a V--A structure
$\bar{N}\gamma_{\mu}\frac{1}{2}(1-\gamma_5)N$. This difference persists
even if the spins of the heavy neutrinos are summed over, in which case
\ba
\sum_{\vec{n},\vec{n}'}(\frac{d\sigma}{d\Omega})_+&=&
2\;\sigma_0\;\beta^3\;\Bigl[\;f_1(1+C^2)\;\Bigr]\;\;,\nn\\
\sum_{\vec{n},\vec{n}'}(\frac{d\sigma}{d\Omega})_-&=&
4\;\sigma_0\;\beta\;\Bigl[\;f_1(2-\beta^2+C^2\beta^2)\;\Bigr]\;\;,\\
\sum_{\vec{n},\vec{n}'}(\frac{d\sigma}{d\Omega})_d&=&
2\;\sigma_0\;\beta\;\Bigl[\;f_1(1+C^2\beta^2)+f_2(2C\beta)\;\Bigr]\;\;.\nn
\ea
Whereas the spin-averaged Majorana cross sections are forward-backward
symmetric, the Dirac process has a term linear in $\cos\theta$, with a
coefficient proportional to $f_2=4\;c_Vc_A$. Eq.~(2.12) also shows that
the threshold behaviour is $\beta^3$, $\beta$ and $\beta$ for the cases
'$+$', '$-$' and 'd' respectively. In the asymptotic limit $\beta\to 1$
the Majorana cases '$+$' and '$-$' have the same angular distribution
$(1+C^2)$, distinct from that of the Dirac process.
\item[(c)] In the high energy limit $\beta\to 1$, the Dirac process
$\E\to N\bar{N}$ has a spin-dependence given by
\ba
(\frac{d\sigma}{d\Omega})_d & = & \frac{1}{2}\sigma_0\;\beta
\;\Bigl[\;1-(n_z+n_z')+n_zn_z'\;\Bigr]\Bigl[\;
f_1\;(\;1+C^2\;)\;+\;2\;f_2\;C\;\Bigr]\;\;. 
\ea 
The fact that only the longitudinal components ($n_z$ and $n_z'$) of
the $N$, $\bar{N}$ spins appear in this expression is consistent with
the expectation that relativistic Dirac neutrinos are eigenstates of
helicity. The fact that the cross section (2.13) vanishes when $n_z=-1$,
$n_z'=+1$ confirms the expectation that for a V--A current the $N$ and 
$\bar{N}$ are produced in left-handed and right-handed states, respectively.
By comparison, the Majorana processes $\E\to N_1N_2$, for $\eta_{CP}=+1$
and $-1$, have the high energy behaviour ($\beta\to1$)
\ba
(\frac{d\sigma}{d\Omega})_+ &=& 
\frac{1}{2}\sigma_0\;
\Bigl\{\;f_1\;\bigl[(1+C^2)(1+n_zn_z')+S^2(n_yn_y'-n_xn_x')\bigr]
\;-\;2f_2C\;(n_z+n_z')\;\Bigr\}\;,\\\nn\\
(\frac{d\sigma}{d\Omega})_- &=& 
\frac{1}{2}\sigma_0\;
\Bigl\{\;f_1\;\bigl[(1+C^2)(1+n_zn_z')+S^2(n_xn_x'-n_yn_y')\bigr]
\;-\;2f_2C\;(n_z+n_z')\;\Bigr\}\;.
\ea
Contrary to the Dirac case, the Majorana reactions have an explicit dependence
on $n_x$, $n_y$ and $n_x'$, $n_y'$, reflecting the fact that a relativistic
Majorana particle with $m_N\neq 0$ is \underline{not} 
necessarily an eigenstate of helicity, 
and can have a spin pointing in an arbitrary direction. The Majorana cases 
'$+$' and '$-$' differ in the sign of the term proportional to $S^2$, which
contains the transverse (x- and y-) components of the neutrino spins.
It is with the purpose of exposing the subtle differences in the spin state
of the $N_1N_2$ and $N\bar{N}$ systems that we investigate in the following
sections the dilepton final state created by the decays of the heavy neutrinos
via $N_{1,2}\to W^{\pm}e^{\mp}$ and $N(\bar{N})\to W^+e^-(W^-e^+)$.
\end{description}

\section{Like-Sign Dileptons: The Reaction $\E\to N_1N_2\to W^+W^+\Emm$}
\setcounter{equation}{0}
As seen in the preceding, the spin state and the angular distribution of the
Majorana pair produced in $\E\to N_1N_2$ depends on the relative $CP$-parity
$\eta_{CP}$ of the two particles. We wish to see how these differences
manifest themselves in the decay products of $N_1$ and $N_2$. To this end, 
we assume that $m_N>m_W$, and that the simplest decay mechanism is
$N_{1,2}\to W^{\mp}e^{\pm}$. In particular, the reaction sequence 
$\E\to N_1N_2\to W^+W^+e^-e^-$ leads to the appearance of two like-sign
leptons in the final state, an unmistakable signature of Majorana pair
production. (For the purpose of this paper we assume that the $W$-bosons
decay into quark jets, thus avoiding the complications of final states with
3 or 4 charged leptons.)

We have calculated the amplitude of the process $\E\to N_1N_2\to W^+W^+e^-e^-$,
depicted in Fig.~2, assuming a decay interaction ($\alpha_N^{'}$ and 
$\alpha_N^{''}$ beeing real parameters)
\ba
{\cal L}_2(x)\;=\;
-\;\frac{g}{\sqrt{2}} & \biggl[ & \alpha_N^{'}\;\bar{e}\;(x)\;\gamma_\mu\;\frac{1}{2}
(1-\gamma_5)\;N_1(x)\;W^{\mu-}(x)\nn\\
& + & \alpha_N^{'}\; \bar{N}_1(x)\;\gamma_\mu\;
\frac{1}{2}(1-\gamma_5)\;e\;(x)\;W^{\mu+}(x)\nn\\
& + & \alpha_N^{''}\;\bar{e}\;(x)\;\gamma_\mu\;\frac{1}{2}
(1-\gamma_5)\;N_2(x)\;W^{\mu-}(x)\nn\\
& + & \alpha_N^{''}\;\bar{N}_2(x)\;\gamma_\mu\;
\frac{1}{2}(1-\gamma_5)\;e\;(x)\;W^{\mu+}(x)\;\;\;\;\biggr]\;\;. 
\ea
This amplitude has the form (see Appendix for details)
\ba
{\cal M}  & = &i A\;\;j_{\mu}^e\;\;\Delta_Z^{\mu\nu}
                \frac{1}{q_1^2-m_1^2+im_1\Gamma_1}\cdot
                \frac{1}{q_2^2-m_2^2+im_2\Gamma_2}\cdot\nn\\
               &\biggl[&m_2\lambda_2 
               \bar{u}_{t_1}(k_1)\;\gamma_{\rho}\;\sla{q}_1\;
               \gamma_{\nu}\;\gamma_{\sigma}\;\frac{1}{2}(1+\gamma_5)
               \;v_{t_2}(k_2)\nn\\
               &-&m_1\lambda_1\bar{u}_{t_2}(k_2)\;\gamma_{\sigma}\;\sla{q}_2\;
               \gamma_{\nu}\;\gamma_{\rho}\;\frac{1}{2}(1+\gamma_5)
               \;v_{t_1}(k_1)\;\biggr]\;
\epsilon_{\lambda_3}^{*\rho}(k_3)\epsilon_{\lambda_4}^{*\sigma}(k_4)\;\;,  
\ea
where 
\be
A\;=\;\alpha_N\;\alpha_N^{'}\;\alpha_N^{''}
\cdot\frac{g^4}{8\cos^2\theta_W}\;\;.
\ee 
Using the narrow-width approximation for the $N_1$, $N_2$ propagators, and 
specializing to the case $m_1=m_2=m_N$, we obtain the following expression
for the squared matrix element (summed over final and averaged over initial
spins), the subscript in ${\cal M}_{\pm}$ denoting $\eta_{CP}=\pm 1$
($q=p_1+p_2$, $l=p_1-p_2$):
\ba
&&\ovl{|{\cal M}_{\pm}}|^2\;\;\;=\;\;\;\frac{|A|^2}{2}\;\frac{1}{(s-m_Z^2)^2}
\;\frac{\pi}{m_N\Gamma_n}\delta(q_1^2-m_N^2)\;\frac{\pi}{m_N\Gamma_n}
\delta(q_2^2-m_N^2)\;\frac{m_N^2}{m_W^4}\cdot\nn\\
&&\biggl\{\;\;\;\;f_1\cdot\biggl(\;\;\mp(m_N^2-m_W^2)^2(m_N^2+2m_W^2)^2\cdot s
\;\mp 4(m_N^2-2m_W^2)^2\cdot\nn\\
&&\;\;\;\;\;\;\;\Bigl[\;s\;(k_1k_2)(q_1q_2)-s\;(k_1q_2)(k_2q_1)
-(k_1k_2)(q_1q)(q_2q)+(k_1k_2)(q_1l)(q_2l)\nn\\
&&\;\;\;\;\;\;\;+\;(k_1q_2)(k_2q)(q_1q)-(k_1q_2)(k_2l)(q_1l)
+(k_1q)(k_2q_1)(q_2q)-(k_1l)(k_2q_1)(q_2l)\nn\\
&&\;\;\;\;\;\;\;-\;(k_1q)(k_2q)(q_1q_2)+(k_1l)(k_2l)(q_1q_2)
\pm m_N^2\;((k_1q)(k_2q)-(k_1l)(k_2l))\;\Bigr]\nn\\
&&\;\;\;\;+\;2\;(m_N^2-m_W^2)(m_N^2-2m_W^2)^2\;\Bigl[\;(k_1q)(q_2q)-(k_1l)
(q_2l)+(k_2q)(q_1q)-(k_2l)(q_1l)\;\Bigr]\nn\\
&&\;\;\;\;+\;8\;m_W^2(m_N^2-m_W^2)^2\;\Bigl[\;(q_1q)(q_2q)-(q_1l)(q_2l)\;\Bigr]
\;\;\biggr)\nn\\
&&-2f_2\cdot(m_N^2-m_W^2)(m_N^4-4m_W^4)\;
\biggl(\;\;\pm(k_1q)(q_1l)-(k_1q)(q_2l)\nn\\
&&\qquad\mp(k_1l)(q_1q)+(k_1l)(q_2q)\pm(k_2q)(q_2l)-(k_2q)(q_1l)\nn\\
&&\qquad\mp(k_2l)(q_2q)+(k_2l)(q_1q)\;\;\biggr)\;\;\;\;\biggr\}\;\;.
\ea
If the final state is $\Epp$ instead of $\Emm$, we replace $f_2\to -f_2$
in the above equation.

The expression for $\ovl{|{\cal M}_{\pm}}|^2$ can be integrated over the 
phase space of $W^+$ and $W^+$ (i.e. over the momenta $k_3(=q_1-k_1)$ and
$k_4(=q_2-k_2)$, in order to obtain the spectra in the lepton variables
$k_1$ and $k_2$. Defining the four-vectors $k_1$ and $k_2$ in the 
$\E$ c.m.~frame by
\ba
k_1^\mu & = &E_1\;(1,\sin\theta_1\cos\phi_1,\sin\theta_1\sin\phi_1,
                      \cos\theta_1)\;\;,\nn\\ 
k_2^\mu & = &E_2\;(1,\sin\theta_2\cos\phi_2,\sin\theta_2\sin\phi_2,
                      \cos\theta_2)\;\;, 
\ea
we have been able to derive the correlated distribution of the energies
$E_1$ and $E_2$, as well as the correlation of the variables $\cos\theta_1$ 
and $\cos\theta_2$ measured relative to the $e^-$ beam direction.
\subsection{Energy Correlation}
The normalized spectrum in the energies of the dilepton pair $\Emm$ is
(${\cal E}_{1,2}= E_{1,2}/m_N$)
\ba
\frac{1}{\sigma}\cdot(\frac{d\sigma}{d{\cal E}_1d{\cal E}_2})\;\;=\;\;
{\cal N}\;[a+b({\cal E}_1+{\cal E}_2)+c({\cal E}_1+{\cal E}_2)^2
-c({\cal E}_1-{\cal E}_2)^2]\;,
\ea
where ${\cal N}$ is a normalization factor,
\be
{\cal N}=[\;{\cal W}^2\beta^2(a+b\cdot {\cal W}+c\cdot {\cal W}^2)\;]^{-1}
\;\;,
\ee
with ${\cal W}=\sqrt{s}\cdot(m_N^2-m_W^2)/2m_N^3$, $\beta=(1-4m_N^2/s)^{1/2}$.
The coefficients $a$, $b$, $c$ depend on the relative $CP$ of the $N_1N_2$ 
system, and take the values
\ba
a^+&=&m_N^2m_W^2(m_N^2-m_W^2)^2(2\;s-\frac{(m_N^2+2m_W^2)^2}{2m_N^2})
\;\;,\nn\\
b^+&=&\sqrt{s}\;m_N^3(m_N^2-2m_W^2)^2(m_N^2-m_W^2)\;\;,
\qquad\qquad\qquad\qquad\qquad\;(\eta_{CP}=+1)\nn\\
c^+&=&-m_N^6(m_N^2-2m_W^2)^2\;\;,\\
a^-&=&2\;m_W^2(m_N^2-m_W^2)^2(s(s-2m_N^2)-
\frac{m_N^2}{m_W^2}\;(m_N^2+2m_W^2)^2)\;\;,\nn\\
b^-&=&\sqrt{s}\;m_N(m_N^2-2m_W^2)^2(m_N^2-m_W^2)(s-2m_N^2)\;\;,
\qquad\qquad\qquad(\eta_{CP}=-1)\nn\\
c^-&=&-m_N^4(m_N^2-2m_W^2)^2(s-2m_N^2)\;\;.
\ea
Notice that the ratios $b^+/a^+$ and $b^-/a^-$ are unequal (likewise the 
ratios $c^+/a^+$ and $c^-/a^-$), although $b^+/c^+=b^-/c^-$. Thus the 
energy correlation of the two electrons in the final state is different
for the cases $\eta_{CP}=\pm 1$. This is illustrated in Fig.~3 for the 
hypothetical parameters $m_N=500$ GeV, $\sqrt{s}=1200$ GeV. It may be
noted that the factor $f_2=4c_Vc_A$ does not appear in the spectrum
($d\sigma/d{\cal E}_1d{\cal E}_2$), so that the energy correlation
of $\Epp$ dileptons is the same as that of $\Emm$. In the limit $\beta\to 1$
the term $a^{\pm}$ dominates and the '$+$' and '$-$' cases are no more
distinguishable.
\subsection{Angular Correlation}
Eq.~(3.4) also allows a calculation of the correlated angular distribution
of the final state $\Emm$ system. Defining the angles $\theta_{1,2}$ as in
Eq.~(3.5), and integrating over all other variables, we find
\ba
&&(\frac{d\sigma}{d\cos\theta_1d\cos\theta_2})^{\pm}\;\;\sim\;\;
\beta\cdot\int\;d\cos{\theta_n}\;\biggl\{\;f_1\cdot\;\Bigl[\;\mp(m_N^2+2m_W^2)^2\;s\cdot
{\cal K}^{\theta_1}_1{\cal K}^{\theta_2}_1\nn\\
&&+\;(m_N^2-2m_W^2)^2\;s\cdot\Bigl(\;\pm{\cal K}^{\theta_1}_2{\cal K}^{\theta_2}_2
(\cos\theta_n\beta-\cos\theta_1)(cos\theta_n\beta+cos\theta_2)\nn\\
&&\;+\;{\cal K}^{\theta_1}_2{\cal K}^{\theta_2}_1
(1+\beta\cos\theta_n\cos\theta_1)+{\cal K}^{\theta_1}_1{\cal K}^{\theta_2}_2
(1-\beta\cos\theta_n\cos\theta_2)\;\Bigr)\nn\\
&&+\;2m_W^2\;s^2\cdot{\cal K}^{\theta_1}_1
{\cal K}^{\theta_2}_1\;(1+\cos^2\theta_n\beta^2)\;-\;4m_N^2(m_N^2-2m_W^2)^2\nn
\\
&&\cdot{\cal K}^{\theta_1}_2{\cal K}^{\theta_2}_2(1-\cos\theta_1\cos\theta_2)
\;\Bigr]\nn\\
&&+f_2\cdot 2(m_N^4-4m_W^4)\;s\cdot
\left[ \begin{array}{r@{\quad:\quad}l} 
\cos\theta_n\beta\;({\cal K}^{\theta_1}_2
{\cal K}^{\theta_2}_1 - {\cal K}^{\theta_1}_1{\cal K}^{\theta_2}_2) & + 
\\{\cal K}^{\theta_1}_2{\cal K}^{\theta_2}_1\cos\theta_1+
{\cal K}^{\theta_1}_1{\cal K}^{\theta_2}_2\cos\theta_2 & -
\end{array} \right]\;\Bigr\}\;\;, 
\ea
with
\ba
{\cal K}^{\theta_{1(2)}}_1&=&\frac{2A_{1(2)}}{(A_{1(2)}^2-B_{1(2)}^2)^{3/2}}
\;\;,\;\;
{\cal K}^{\theta_{1(2)}}_2\;\;=\;\;\frac{2A_{1(2)}^2+B_{1(2)}^2}{(A_{1(2)}^2-B_{1(2)}^2)^{5/2}}\;\;,\\\nn\\
A_{1(2)}&=&1-(+)\beta\cos\theta_n\cos\theta_{1(2)}\;\;,\;\;
B_{1(2)}\;\;=\;\;\beta\sin\theta_n\sin\theta_{1(2)}\;\;.\nn
\ea
The correlation (3.10) has been evaluated for $m_N=500$ GeV and 
$\sqrt{s}=1200$ GeV (using $f_1=1+4\sin^2\theta_W+8\sin^4\theta_4$,
$f_2=1-4\sin^2\theta_W$) and is plotted in Fig.~4. There is a clear difference
between the cases $\eta_{CP}=\pm 1$. The angular correlation in (3.10) becomes
particularly transparent near the threshold $\beta\to 0$, where we obtain the 
analytic results
\ba
\frac{1}{\sigma^+}\cdot
(\frac{d\sigma}{d\cos\theta_1d\cos\theta_2})_{\beta\to 0}^+
&\approx&\frac{1}{4}\cdot[\;1+\frac{1}{2}\cdot\frac{f_2}{f_1}\cdot
\frac{m_N^2-2m_W^2}{m_N^2+2m_W^2}\cdot
(\cos\theta_1+\cos\theta_2)\;]\;,\\\nn\\
\frac{1}{\sigma^-}\cdot
(\frac{d\sigma}{d\cos\theta_1d\cos\theta_2})_{\beta\to 0}^-
&\approx&\frac{1}{4}\cdot[\;1+\frac{(m_N^2-2m_W^2)^2}{(m_N^2+2m_W^2)^2}
\cdot\cos\theta_1\cos\theta_2\nn\\
&&\quad\;+\frac{f_2}{f_1}\cdot\frac{(m_N^2-2m_W^2)}{(m_N^2+2m_W^2)}
\cdot(\cos\theta_1+\cos\theta_2)\;]\;.
\ea
Notice that the distribution in the variables $\cos\theta_1$ and $\cos\theta_2$
becomes flat in the case $\eta_{CP}=+1$ when $f_2/f_1$ is neglected. 
By contrast, there remains a nontrivial correlation for $\eta_{CP}=-1$, even
in the absence of $f_2$. As before, the above results for $\Emm$ hold for
$\Epp$ if one replaces $f_2\to-f_2$.

\section{Unlike-Sign Dileptons: The Reaction $\E\to N_1N_2\to W^+W^-\E$}
\setcounter{equation}{0}
Proceeding as in Sec.~3, the matrix element for the reaction
$\E\to N_1N_2\to W^+W^-\E$ (Fig.~2) is
\ba
{\cal M}_m & = &i A\;\;j_{\mu}^e\;\;\Delta_Z^{\mu\nu}
                \frac{1}{q_1^2-m_1^2+im_1\Gamma_1}\cdot
                \frac{1}{q_2^2-m_2^2+im_2\Gamma_2}\cdot\nn\\
               &[&\lambda_2\; 
               \bar{u}_{t_1}(k_1)\;\gamma_{\rho}\;\sla{q}_1\;\gamma_{\nu}\;
               \sla{q}_2\;\gamma_{\sigma}\;\frac{1}{2}(1-\gamma_5)
               \;v_{t_2}(k_2)\\
               &+&\lambda_1\;m_1m_2\;\bar{u}_{t_1}(k_1)\;
               \gamma_{\rho}\;
               \gamma_{\nu}\;\gamma_{\sigma}\;\frac{1}{2}(1-\gamma_5)
               \;v_{t_2}(k_2)\;]\;
\epsilon_{\lambda_3}^{*\rho}(k_3)\epsilon_{\lambda_4}^{*\sigma}(k_4)\;\;.\nn  
\ea
The same final state, produced via a Dirac particle-antiparticle pair
($\E\to N\bar{N}\to W^+W^-\E$),  has the amplitude
\ba 
{\cal M}_d & = &i A\;\;j_{\mu}^e\;\;\Delta_Z^{\mu\nu}
                \frac{1}{q_1^2-m_1^2+im_1\Gamma_1}\cdot
                \frac{1}{q_2^2-m_2^2+im_2\Gamma_2}\cdot\nn\\
&&\bar{u}_{t_1}(k_1)\;\gamma_{\rho}\;\sla{q}_1\;\gamma_{\nu}\;
              \sla{q}_2\;\gamma_{\sigma}\;\frac{1}{2}(1-\gamma_5)
              \;v_{t_2}(q_2)\;\epsilon_{\lambda_3}^{*\rho}(k_3)
              \epsilon_{\lambda_4}^{*\sigma}(k_4)\;\;.
\ea
Summing (averaging) over final (initial) polarizations, and using the 
narrow-width approximation for the $N_1$, $N_2$ propagators, we obain the 
squared matrix elements given below:
\ba
&&\ovl{|{\cal M}_{\pm}|}^2\;\;\;=\;\;\;\frac{|A|^2}{2}\;\frac{1}{(s-m_Z^2)^2}
\;\frac{\pi}{m_N\Gamma_n}\delta(q_1^2-m_N^2)\;\frac{\pi}{m_N\Gamma_n}
\delta(q_2^2-m_N^2)\;
\frac{m_N^2}{m_W^4}\cdot\nn\\
&&\biggl\{\;\;\;\;f_1\cdot\biggl(\;\;\mp(m_N^2-m_W^2)^2(m_N^2+2m_W^2)^2\cdot s
\;\pm 4(m_N^2-2m_W^2)^2\cdot\nn\\
&&\;\;\;\;\;\;\;\Bigl[\;s\;(k_1k_2)(q_1q_2)-s\;(k_1q_2)(k_2q_1)
-(k_1k_2)(q_1q)(q_2q)+(k_1k_2)(q_1l)(q_2l)\nn\\
&&\;\;\;\;\;\;\;+\;(k_1q_2)(k_2q)(q_1q)-(k_1q_2)(k_2l)(q_1l)
+(k_1q)(k_2q_1)(q_2q)-(k_1l)(k_2q_1)(q_2l)\nn\\
&&\;\;\;\;\;\;\;-\;(k_1q)(k_2q)(q_1q_2)+(k_1l)(k_2l)(q_1q_2)
\pm m_N^2\;((k_1q)(k_2q)-(k_1l)(k_2l))\;\Bigr]\nn\\
&&\;\;\;\;-\;2\;(m_N^2-m_W^2)(m_N^2-2m_W^2)^2\;\Bigl[\;(k_1q)(q_2q)-(k_1l)
(q_2l)+(k_2q)(q_1q)-(k_2l)(q_1l)\;\Bigr]\nn\\
&&\;\;\;\;+\;2\;(m_N^2+4m_W^2/m_N^2)(m_N^2-m_W^2)^2\;
[\;(q_1q)(q_2q)-(q_1l)(q_2l)\;]
\;\;\biggr)\nn\\
&&-2f_2\cdot\biggl(\;(m_N^2-m_W^2)(m_N^4-4m_W^4)\;(\;\;\pm (k_1q)(q_1l)-(k_1q)(q_2l)
\nn\\
&&\qquad\mp(k_1l)(q_1q)+(k_1l)(q_2q)\mp(k_2q)(q_2l)+(k_2q)(q_1l)\nn\\
&&\qquad\pm(k_2l)(q_2q)-(k_2l)(q_1q)\;\;)\nn\\
&&\;\;\;\;\;\;\;+2(m_N^2+4m_W^4/m_N^2)(m_N^2-m_W^2)^2\;
\Bigl[(q_1q)(q_2l)-(q_2q)(q_1l)\Bigr]
\;\biggr)\;\;\;\;\biggr\}\;\;,
\ea
\ba
&&\ovl{|{\cal M}_d|}^2\;\;\;=\;\;\;|A|^2\;\frac{1}{(s-m_Z^2)^2}
\;\frac{\pi}{m_N\Gamma_n}\delta(q_1^2-m_N^2)\;\frac{\pi}{m_N\Gamma_n}
\delta(q_2^2-m_N^2)\nn\\
&\biggl\{&f_1\cdot\biggl(\;\;\frac{m_N^4}{m_W^4}(m_N^2-2m_W^2)^2\cdot
\Bigl[(k_1q)(k_2q)-(k_1l)(k_2l)\Bigr]\nn\\
&&+2\frac{m_N^2}{m_W^2}(m_N^2-m_W^2)(m_N^2-2m_W^2)
\cdot\Bigl[(k_1q)(q_2q)-(k_1l)(q_2l)+(k_2q)(q_1q)-(k_2l)(q_1l)\Bigr]\nn\\
&&+4(m_N^2-m_W^2)^2\cdot\Bigl[(q_1q)(q_2q)-(q_1l)(q_2l)\Bigr]\;\;\biggr)
\nn\\
&+&f_2\cdot\biggl(\;\;\frac{m_N^4}{m_W^4}(m_N^2-2m_W^2)^2\cdot\Bigl
[(q_1q)(k_2l)-(k_2q)(k_1l)\Bigr]\nn\\
&&+2\frac{m_N^2}{m_W^2}(m_N^2-m_W^2)(m_N^2-2m_W^2)
\cdot\Bigl[(k_1q)(q_2l)-(k_2q)(q_1l)+(q_1q)(k_2l)-(q_2q)(k_1l)\Bigr]\nn\\
&&+4(m_N^2-m_W^2)^2\cdot\Bigl[(q_1q)(q_2l)-(q_2q)(q_1l)\Bigr]\;\;\biggr)
\;\;\;\biggr\}\;\;.
\ea
In complete analogy with the discussion of like-sign leptons (Sec.~3), we
derive from the above equations the correlation in the energies and angles 
of the final $\E$ state.
\subsection{Energy Correlation}
The distribution in the scaled energies ${\cal E}_1$, ${\cal E}_2$ has the
quadratic form given in Eq.~(3.6), where the coefficients in the 
Majorana cases '$+$' and '$-$' and the Dirac case 'd' now have the values
\ba
a^+&=&\frac{1}{2}m_N^2(m_N^2-m_W^2)^2\biggl(\frac{s}{m_N^2}(m_N^4+4m_W^4)-
2(m_N^2+2m_W^2)^2\biggr)\;\;,\nn\\
b^+&=&-\sqrt{s}\;m_N^3(m_N^2-2m_W^2)^2(m_N^2-m_W^2)\;\;,\nn\\
c^+&=&m_N^6(m_N^2-2m_W^2)^2\;\;,\nn\\
a^-&=&\frac{1}{2}m_N^2(m_N^2-m_W^2)^2\biggl(s(s-2m_N^2)(1+4\frac{m_W^4}{m_N^4})
-4(m_N^2+2m_W^2)^2\biggr)\;\;,\nn\\
b^-&=&-\sqrt{s}\;m_N(m_N^2-2m_W^2)^2(m_N^2-m_W^2)(s-2m_N^2)\;\;,\nn\\
c^-&=&m_N^4(m_N^2-2m_W^2)^2(s-2m_N^2)\;\;,\nn\\
a^d&=&(m_N^2-m_W^2)^2\biggl(4(s-m_N^2)(s-4m_N^2)\frac{m_W^4}{m_N^2}\nn\\
   &&-(m_N^2-2m_W^2)
\Bigl(2(s-4m_N^2)m_W^2-m_N^2(m_N^2-2m_W^2)\Bigr)\biggr)\;\;,\nn\\
b^d&=&\sqrt{s}\;m_N(m_N^2-2m_W^2)(m_N^2-m_W^2)\biggl(4sm_W^2-(m_N^2+14m_W^2)
m_N^2\biggr)\;\;,\nn\\
c^d&=&m_N^4(m_N^2-2m_W^2)^2(s-3m_N^2)\;\;.
\ea
The corresponding three distributions are plotted in Fig.~5. As in the case of
like-sign dileptons, the $\E$ pairs have distinct correlations for 
$\eta_{CP}=\pm 1$. A comparison of the Majorana cases with the Dirac case
reveals an interesting difference. In the Majorana cases the total $\E$
energy $Y={\cal E}_1+{\cal E}_2$ is distributed symmetrically around the
mid-point of this variable $Y_0=1/2(Y_{min}+Y_{max})$. By contrast,
$\E$ pairs resulting from Dirac $N\bar{N}$ primary state have a total energy 
distribution that is unsymmetric around the mid-point. 
\subsection{Angle Correlation}
In analogy to the distribution $d\sigma/d\cos\theta_1d\cos\theta_2$
obtained for $\Emm$ pairs (Eq.~3.10), 
the result for unlike-sign dileptons $\E$ is
\ba
&&(\frac{d\sigma}{d\cos\theta_1d\cos\theta_2})^{\pm}\;\;\sim\;\;
\beta\cdot\int\;d\cos{\theta_n}\;\biggl\{\;f_1\cdot\nn\\\;
&&\Bigl[\;\mp 2m_N^2(m_N^2+2m_W^2)^2\;s\cdot
{\cal K}^{\theta_1}_1{\cal K}^{\theta_2}_1\nn\\
&&-\;2m_N^2(m_N^2-2m_W^2)^2\;s\cdot\Bigl(\;\pm{\cal K}^{\theta_1}_2{\cal K}^{\theta_2}_2
(\cos\theta_n\beta-\cos\theta_1)(cos\theta_n\beta+cos\theta_2)\nn\\
&&\;+\;{\cal K}^{\theta_1}_2{\cal K}^{\theta_2}_1
(1+\beta\cos\theta_n\cos\theta_1)+{\cal K}^{\theta_1}_1{\cal K}^{\theta_2}_2
(1-\beta\cos\theta_n\cos\theta_2)\;\Bigr)\nn\\
&&+\;(m_N^4+4m_W^4)\;s^2\cdot{\cal K}^{\theta_1}_1
{\cal K}^{\theta_2}_1\;(1+\cos^2\theta_n\beta^2)\;+\;8m_N^4(m_N^2-2m_W^2)^2
\cdot\nn\\
&&\;\;\;{\cal K}^{\theta_1}_2{\cal K}^{\theta_2}_2(1-\cos\theta_1\cos\theta_2)
\;\Bigr]\nn\\
&&+f_2\cdot 2(m_N^4-4m_W^4)\;s\cdot
\Bigl[\;-{\cal K}^{\theta_1}_1{\cal K}^{\theta_2}_1\cdot s\;\cos\theta_n\beta
\nn\\
&&\;\;\;+2m_N^2\cdot\left[ \begin{array}{r@{\quad:\quad}l} 
\cos\theta_n\beta\;({\cal K}^{\theta_1}_2
{\cal K}^{\theta_2}_1 + {\cal K}^{\theta_1}_1{\cal K}^{\theta_2}_2) & + 
\\{\cal K}^{\theta_1}_2{\cal K}^{\theta_2}_1\cos\theta_1-
{\cal K}^{\theta_1}_1{\cal K}^{\theta_2}_2\cos\theta_2 & -
\end{array} \right]\;\Bigr]\;\;\biggr\}\;\;,\\\nn\\ 
&&(\frac{d\sigma}{d\cos\theta_1d\cos\theta_2})^{d}\;\;\sim\;\;
\beta\cdot\int\;d\cos{\theta_n}\;\biggl\{\;f_1\cdot\nn\\\;
&&\Bigl[\;s^2m_W^4\cdot{\cal K}^{\theta_1}_1{\cal K}^{\theta_2}_1\cdot
(1+\cos\theta_n\beta^2)\nn\\
&&+\;m_N^2m_W^2(m_N^2-2m_W^2)\;s\cdot\Bigl(\;{\cal K}^{\theta_1}_2
{\cal K}^{\theta_2}_2(1+\beta\cos\theta_n\cos\theta_1)+
{\cal K}^{\theta_1}_1{\cal K}^{\theta_2}_2
(1-\beta\cos\theta_n\cos\theta_2)\;\Bigr)\nn\\
&&+\;m_N^4(m_N^2-2m_W^2)^2\cdot{\cal K}^{\theta_1}_2{\cal K}^{\theta_2}_2\cdot
(1-\cos\theta_1\cos\theta_2)\;\Bigr]
\nn\\
&&+f_2\cdot\Bigl[\;2s^2m_W^4\cdot
{\cal K}^{\theta_1}_1{\cal K}^{\theta_2}_1\cdot
\cos\theta_n\beta\nn\\
&&+\;m_N^2m_W^2(m_N^2-2m_W^2)\;s\cdot\Bigl(\;{\cal K}^{\theta_1}_2
{\cal K}^{\theta_2}_2(\cos\theta_n\beta+\cos\theta_1)+
{\cal K}^{\theta_1}_1{\cal K}^{\theta_2}_2
(\cos\theta_n\beta-\cos\theta_2)\;\Bigr)\nn\\
&&+\;m_N^4(m_N^2-2m_W^2)^2\cdot{\cal K}^{\theta_1}_2{\cal K}^{\theta_2}_2\cdot
(\cos\theta_1-\cos\theta_2)\;\Bigr]\;\;\biggr\}\;\;.
\ea
As usual, the indices '$+$', '$-$' and 'd' differentiate between the
Majorana cases $\eta_{CP}=+1$, $-1$ and the Dirac case. The angle-correlations
expressed by Eqs.~(4.6), (4.7) are plotted in Fig.~6., where the differences
between the three cases are obvious. Close to threshold ($\beta\to 0$),
the correlation between $\cos\theta_1$ and $\cos\theta_2$ can be presented
in analytic form
\ba
&&\frac{1}{\sigma^+}\cdot
(\frac{d\sigma}{d\cos\theta_1d\cos\theta_2})_{\beta\to 0}^+
\approx\frac{1}{4}\cdot[\;1+\frac{1}{2}\cdot\frac{f_2}{f_1}\cdot
\frac{m_N^2-2m_W^2}{m_N^2+2m_W^2}\cdot
(\cos\theta_1-\cos\theta_2)\;]\;,\\\nn\\
&&\frac{1}{\sigma^d}\cdot(\frac{d\sigma}{d\cos\theta_1d\cos\theta_2})_{\beta\to 0}^d
=\frac{1}{\sigma^-}\cdot(\frac{d\sigma}{d\cos\theta_1d\cos\theta_2})_{\beta\to 0}^-
\\\nn\\
&&\approx\frac{1}{4}\cdot[\;1-\frac{(m_N^2-2m_W^2)^2}{(m_N^2+2m_W^2)^2}
\cdot\cos\theta_1\cos\theta_2
\;+\frac{f_2}{f_1}\cdot\frac{(m_N^2-2m_W^2)}{(m_N^2+2m_W^2)}
\cdot(\cos\theta_1-\cos\theta_2)\;]\;.\nn
\ea
In this limit, the cases '$+$' and '$-$' remain distinct, but the case
$\eta_{CP}=-1$ converges to the Dirac case.

\section{Comments}
We comment briefly on some other papers which have a partial overlap with
the considerations presented above.
\begin{description}
\item[(i)] Our discussion of the production reaction $\E\to N_1N_2$ follows
very closely that given in Ref.~\cite{Tsai}. Our results for
$d\sigma/d\Omega$ given in the Appendix (Eqs.~(A.1)--(A.6)) 
essentially coincide
with those in this paper, with two minor differences:
The angular distributions for the case of two distinct Majorana particles
with the same $CP$-parity, as well as for the case of two distinct
Dirac particles (Eq.~(4E) and (4D) in Ref.~\cite{Tsai}), are slightly
different from our distributions, presented in the Appendix (Eq.~(A.2) 
and (A.3)).
\item[(ii)] The cross section for the Majorana process $\E\to N_1N_2$, 
with $m_1=m_2$ and $\eta_{CP}=+1$ calculated in Ref.~\cite{Ma} agrees with 
that obtained in this paper. However the Dirac case $\E\to N\bar{N}$
(Eq.~(2) of Ref.~\cite{Ma}) differs from our result (Eq.~(A.5)), as also 
noted in Ref.~\cite{Tsai}.
\item[(iii)] The spin-summed differential cross section for the Majorana
process $\E\to N_1N_2$ (with $m_1=m_2$, $\eta_{CP}=+1$) calculated in
the present paper, as well as in Refs.~\cite{Tsai,Ma}, differs from that
given in Ref.~\cite{Lan}, but agrees with the results given in 
Refs.~\cite{Maal,Buch,Zer}.
\item[(iv)] Our analysis of heavy Majorana production and decay has been
essentially model-independent. Discussions in the context of specific
gauge models, based on $SU(2)_L\times SU(2)_R\times U(1)$ or $E(6)$
symmetries, may be found in Refs.~\cite{Buch,Zer,Gluza}.  
\end{description}
\newpage

\appendix
\renewcommand{\theequation}{A.\arabic{equation}}
\section{Differential Cross Section for $\E\to N_1N_2$}
Following Ref.~\cite{Tsai}, we consider the following five cases, in which
$N_1$ and $N_2$ are
\begin{description}
\item[A] distinct Dirac particles.
\item[B] distinct Majorana particles with the same $CP$-parity.
\item[C] distinct Majorana particles with opposite $CP$-parity.
\item[D] Dirac particle-antiparticle pair.
\item[E] identical Majorana particles.
\end{description}
Choosing the $N_1$ direction in the $\E$ c.m. system to be the z-axis, and
the $e^-$-beam direction to be at an angle $\theta$ ($=$ scattering angle),
the momenta ($q_1$, $q_2$) and spins ($t_1$, $t_2$) of $N_1$ and $N_2$
have components
\ba
N_1\;:\; q_1^\mu & = & (\gamma m_1,0,0,\gamma\beta m_1)\;\;,\nn\\
         t_1^\mu & = & (\gamma\beta n_z,n_x,n_y,\gamma n_z)
\;\;,\nn\\
N_2\;:\; q_2^\mu & = & (\gamma' m_2,0,0,-\gamma'\beta' m_2)\;\;,\nn\\
         t_2^\mu & = & (-\gamma'\beta' n_z',n_x',n_y',\gamma' n_z')
\;\;.
\ea
The differential cross sections are (with $\beta=(1-4m_1^2/s)^{1/2}$,
$\beta'=(1-4m_2^2/s)^{1/2}$, $\gamma=(1-\beta^2)^{-1/2}$, 
$\gamma'=(1-\beta'^2)^{-1/2}$,
$\lambda(x,y,z)=[\;x^2+y^2+z^2-2(xy+yz+zx)\;]^{1/2}$)
\ba 
(\frac{d\sigma}{d\Omega})_A& = &\frac{G_F^2 \alpha_N^2}{512 \pi^2}
\;|R(s)|^2\;[1-(m_1^2-m_2^2)^2/s^2]\;\lambda(s,m_1^2,m_2^2)\nn\\
&\biggl\{ &f_1\;\;\bigl[\;(1+C^2\beta'\beta)-(\beta n_z'+\beta'n_z)
(1+C^2)-(\beta' n_x/\gamma+\beta n_x'/\gamma')
SC\nn\\
&   &\;\;\;\;\;+\;n_zn_z'(C^2+\beta\beta')+(n_xn_z'/\gamma+
n_x'n_z/\gamma')SC+n_xn_x'S^2/\gamma\gamma'\;\bigr]\nn\\
& + &f_2\;\;\bigl[\;C(\beta+\beta')-(n_z+n_z')C(1+\beta\beta')
-(n_x/\gamma+n_x'/\gamma')S\nn\\
& + &n_zn_z'C(\beta+\beta')+S(\beta' n_xn_z'/\gamma+\beta n_x'n_z/\gamma')
\;\bigr]\;\;\;\biggr\}\;,\;\\\nn\\
(\frac{d\sigma}{d\Omega})_B& = &\frac{G_F^2 \alpha_N^2}{256 \pi^2}
\;|R(s)|^2\;[1-(m_1^2-m_2^2)^2/s^2]\;\lambda(s,m_1^2,m_2^2)\nn\\
& \biggl\{ &f_1\;\;\bigl[\;n_xn_x'S^2(1/\gamma\gamma'-1)+n_yn_y'S^2\beta\beta'
+n_zn_z'(\beta\beta'-C^2(1/\gamma\gamma'-1))\nn\\
&   &\;\;\;\;\;+\;(n_xn_z'-n_x'n_z)SC(1/\gamma-1/\gamma')
+\;C^2\beta\beta'-1/\gamma\gamma'+1\;\bigr]\nn\\
& + &f_2\;\;\bigl[\;(n_x-n_x')S(1/\gamma'-1/\gamma)+(n_z+n_z')C
(1/\gamma\gamma'-\beta\beta'-1)\;\bigr]\;\;\biggr\}\;,\;\\\nn\\
(\frac{d\sigma}{d\Omega})_C& = &\frac{G_F^2 \alpha_N^2}{256 \pi^2}
\;|R(s)|^2\;[1-(m_1^2-m_2^2)^2/s^2]\;\lambda(s,m_1^2,m_2^2)\nn\\
& \biggl\{ &f_1\;\;\bigl[\;n_xn_x'S^2(1/\gamma\gamma'+1)-n_yn_y'S^2\beta\beta'
+n_zn_z'(\beta\beta'+C^2(1/\gamma\gamma'+1))\nn\\
&   &\;\;\;\;\;+\;(n_xn_z'+n_x'n_z)SC(1/\gamma+1/\gamma')
+\;C^2\beta\beta'+1/\gamma\gamma'+1\;\bigr]\nn\\
& - &f_2\;\;\bigl[\;(n_x+n_x')S(1/\gamma+1/\gamma')+(n_z+n_z')C
(1/\gamma\gamma'+\beta\beta'+1)\;\bigr]\;\;\biggr\}\;,\;\\\nn\\
(\frac{d\sigma}{d\Omega})_D & = & 
(\frac{d\sigma}{d\Omega})_{A,m_1=m_2} 
\;\;=\;\; \frac{G_F^2 \alpha_N^2}{512 \pi^2}
\;|R(s)|^2\;\lambda(s,m^2,m^2)\nn\\
& \biggl\{ & f_1\;\;[(1+C^2\beta^2)-(n_z+n_z')\beta(1+C^2)-(n_x+n_x')SC\beta/\gamma
+n_zn_z'(C^2+\beta^2)\nn\\
& + & (n_xn_z'+n_zn_x')SC/\gamma+n_xn_x'S^2/\gamma^2]\;\;+\;\; 
f_2\;\;[\;2C\beta-(n_z+n_z')C(1+\beta^2)\nn\\
& - &(n_x+n_x')S/\gamma+
2n_zn_z'C\beta+(n_xn_z'+n_x'n_z)S\beta/\gamma\;]\;\;\biggr\}\;\;,\\\nn\\
(\frac{d\sigma}{d\Omega})_E & = & 
\frac{1}{2}\cdot(\frac{d\sigma}{d\Omega})_{B,m_1=m_2} 
\;\;=\;\;\frac{G_F^2 \alpha_N^2}{512 \pi^2}
\;|R(s)|^2\;\lambda(s,m^2,m^2)\beta^2\nn\\
& \biggl\{ &f_1\;\;[\;(n_yn_y'-n_xn_x')S^2+(1+n_zn_z')(1+C^2)\;]
\;-\;f_2\;\;2(n_z+n_z')C\;\;\biggr\}\;\;.
\ea

\section{Matrix Elements for $\E\to N_1N_2\to \Eetc$}
\setcounter{equation}{0}
\renewcommand{\theequation}{B.\arabic{equation}}
\subsection{Like Sign Dileptons}
The matrix element for the reaction (Fig.~2)
\be
e^+(p_2,t_2)+e^-(p_1,t_1)\;\to\; e^-(k_1)e^-(k_2)W^+(k_3,\lambda_3)
W^+(k_4,\lambda_4) 
\ee is
\ba
{\cal M}\;=\;i A\;j_{\mu}^e\;\Delta_Z^{\mu\nu}\;\biggl\{&&
\;\lambda_2\;\bar{u}_{t_1}(k_1)\;\gamma_{\rho}\;\frac{1}{2}(1-\gamma_5)
\;\frac{\sla{q}_1+m_1}{q_1^2-m_1^2+im_1\Gamma_1}\;\gamma_{\nu}\;\frac{1}{2}
(1-\gamma_5)\nn\\
&&\quad\;\frac{-\sla{q}_2+m_2}{q_2^2-m_2^2+im_2\Gamma_2}\;\gamma_{\sigma}\;
\frac{1}{2}(1+\gamma_5)\;v_{t_2}(k_2)\nn\\
-&&\;\lambda_1\;\bar{u}_{t_2}(k_2)\;\gamma_{\sigma}\;\frac{1}{2}(1-\gamma_5)
\;\frac{\sla{q}_2+m_2}{q_2^2-m_2^2+im_2\Gamma_2}\;\gamma_{\nu}\;\frac{1}{2}
(1-\gamma_5)\nn\\
&&\quad\;\frac{-\sla{q}_1+m_1}{q_1^2-m_1^2+im_1\Gamma_1}\;\gamma_{\rho}\;
\frac{1}{2}(1+\gamma_5)\;v_{t_1}(k_1)\;\biggr\}\;
\epsilon_{\lambda_3}^{*\rho}(k_3)\epsilon_{\lambda_4}^{*\sigma}(k_4)\;\;.
\ea
Upon rearrangement, this gives the matrix element in Eq.~(3.2).
\subsection{Unlike Sign Dileptons}
The matrix element for the Majorana-mediated process (Fig.~2)
\be
e^+(p_2,t_2)+e^-(p_1,t_1)\;\to\; e^-(k_1)e^-(k_2)W^+(k_3,\lambda_3)
W^+(k_4,\lambda_4) 
\ee is
\ba
{\cal M}_m\;=\;-i A\;j_{\mu}^e\;\Delta_Z^{\mu\nu}\;\biggl\{&&
\;\lambda_2\;\bar{u}_{t_1}(k_1)\;\gamma_{\rho}\;\frac{1}{2}(1-\gamma_5)
\;\frac{\sla{q}_1+m_1}{q_1^2-m_1^2+im_1\Gamma_1}\;\gamma_{\nu}\;\frac{1}{2}
(1-\gamma_5)\nn\\
&&\quad\;\frac{-\sla{q}_2+m_2}{q_2^2-m_2^2+im_2\Gamma_2}\;\gamma_{\sigma}\;
\frac{1}{2}(1-\gamma_5)\;v_{t_2}(k_2)\nn\\
-&&\;\lambda_1\;\bar{u}_{t_1}(k_1)\;\gamma_{\rho}\;\frac{1}{2}(1-\gamma_5)
\;\frac{\sla{q}_1+m_1}{q_1^2-m_1^2+im_1\Gamma_1}\;\gamma_{\nu}\;\frac{1}{2}
(1+\gamma_5)\nn\\
&&\quad\;\frac{-\sla{q}_2+m_2}{q_2^2-m_2^2+im_2\Gamma_2}\;\gamma_{\sigma}\;
\frac{1}{2}(1-\gamma_5)\;v_{t_2}(k_2)\;\biggr\}\;
\epsilon_{\lambda_3}^{*\rho}(k_3)\epsilon_{\lambda_4}^{*\sigma}(k_4)\;\;.
\ea
Upon rearrangement, this gives the matrix element in Eq.~(4.1).

\newpage
\begin{center}
{\large FIGURE CAPTIONS}
\end{center}
\begin{itemize}
\item[1.] Feynman diagram for the reaction $\E\to N_1N_2$
\item[2.] Diagram showing the sequential process $\E\to N_1N_2\to \Eetc$
\item[3.] Energy correlation of the $\Emm$ lepton pair in the reaction
$\E\to N_1N_2\to e^-W^+e^-W^+$, for the cases (a) $\eta_{CP}=+1$, 
(b) $\eta_{CP}=-1$. (Paramers for this and succeeding figures: 
$\sqrt{s}=1.2$ TeV, $m_N=500$ GeV.)
\item[4.] Angle correlation of $\Emm$ dileptons in
$\E\to N_1N_2\to e^-W^+e^-W^+$, for (a) $\eta_{CP}=+1$, (b) $\eta_{CP}=-1$
\item[5.] Energy correlation of $e^-e^+$ dileptons in 
$\E\to N_1N_2\to e^-W^+e^+W^-$: (a) Majorana pair, $\eta_{CP}=+1$,
(b) Majorana pair, $\eta_{CP}=-1$, (c) Dirac $N\bar{N}$-pair
\item[6.] Angle correlation of $e^-e^+$ dileptons in
$\E\to N_1N_2\to e^-W^+e^+W^-$: (a) Majorana pair, $\eta_{CP}=+1$,
(b) Majorana pair, $\eta_{CP}=-1$, (c) Dirac $N\bar{N}$-pair   
\end{itemize}
\newpage    
\begin{figure}[H]
\begin{center}
\mbox{\epsfysize 6cm \epsffile{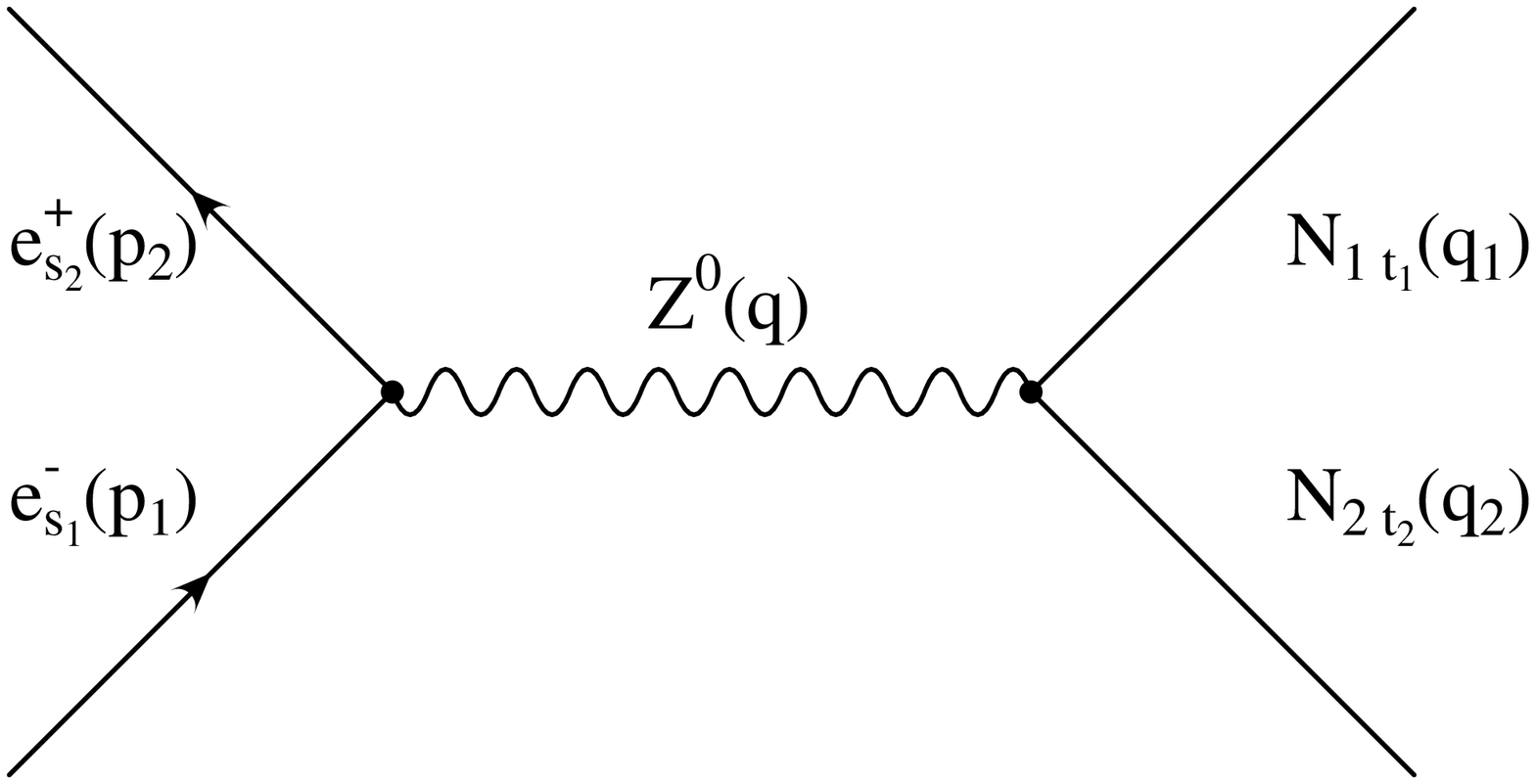}}
\vspace{1.5cm}
\caption{}
\end{center}
\end{figure}
\vspace{1cm}
\begin{figure}[H]
\begin{center}
\vspace{-2cm}
\mbox{\epsfysize 8cm \epsfxsize 10cm \epsffile{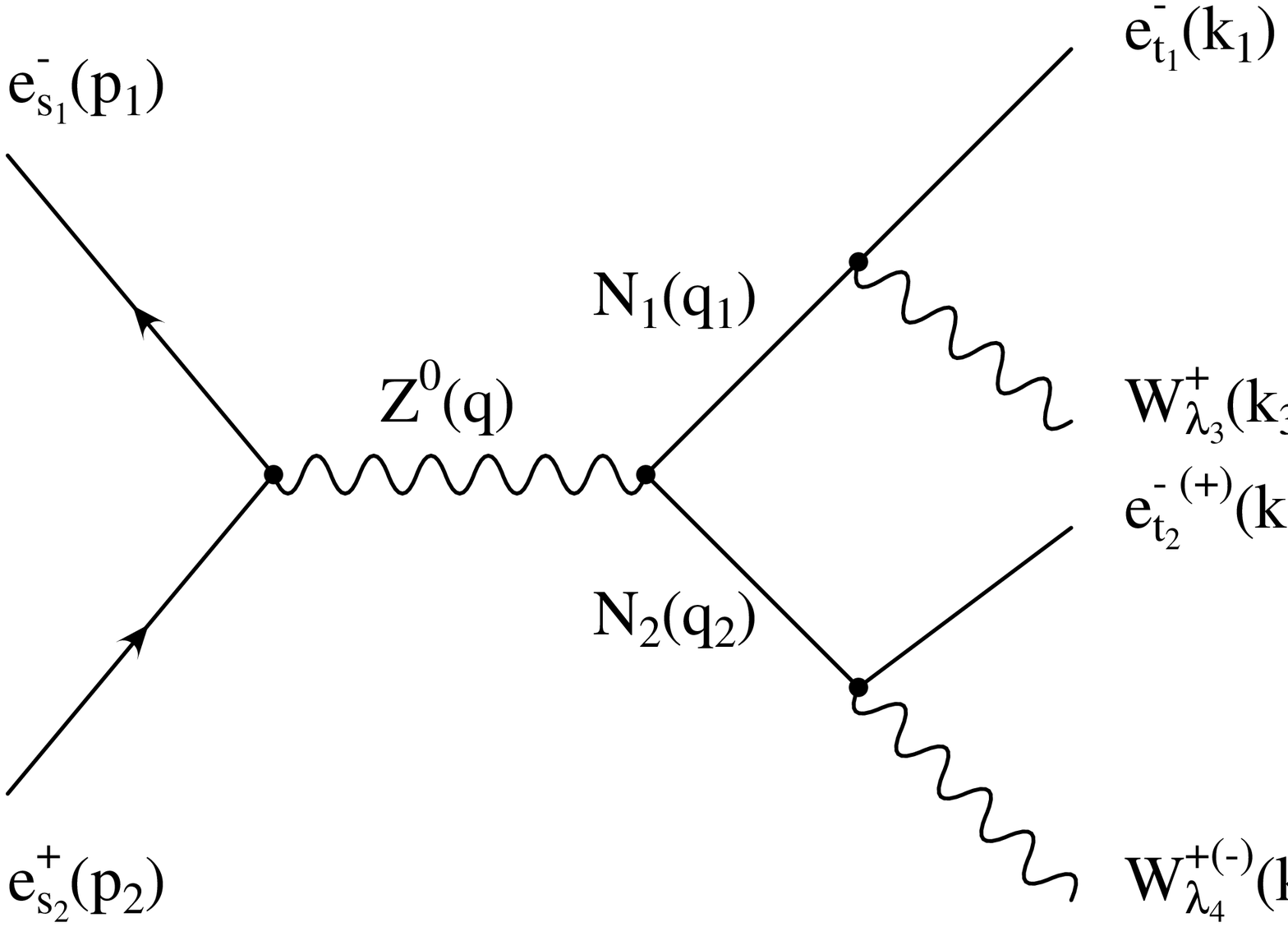}}
\vspace{1.7cm}
\caption{}
\end{center}
\end{figure}
\newpage
\begin{center}
{\Large $\Emm$ Final State: Energy Correlation
\begin{figure}[H]
\mbox{\epsfysize 9.5cm \epsffile{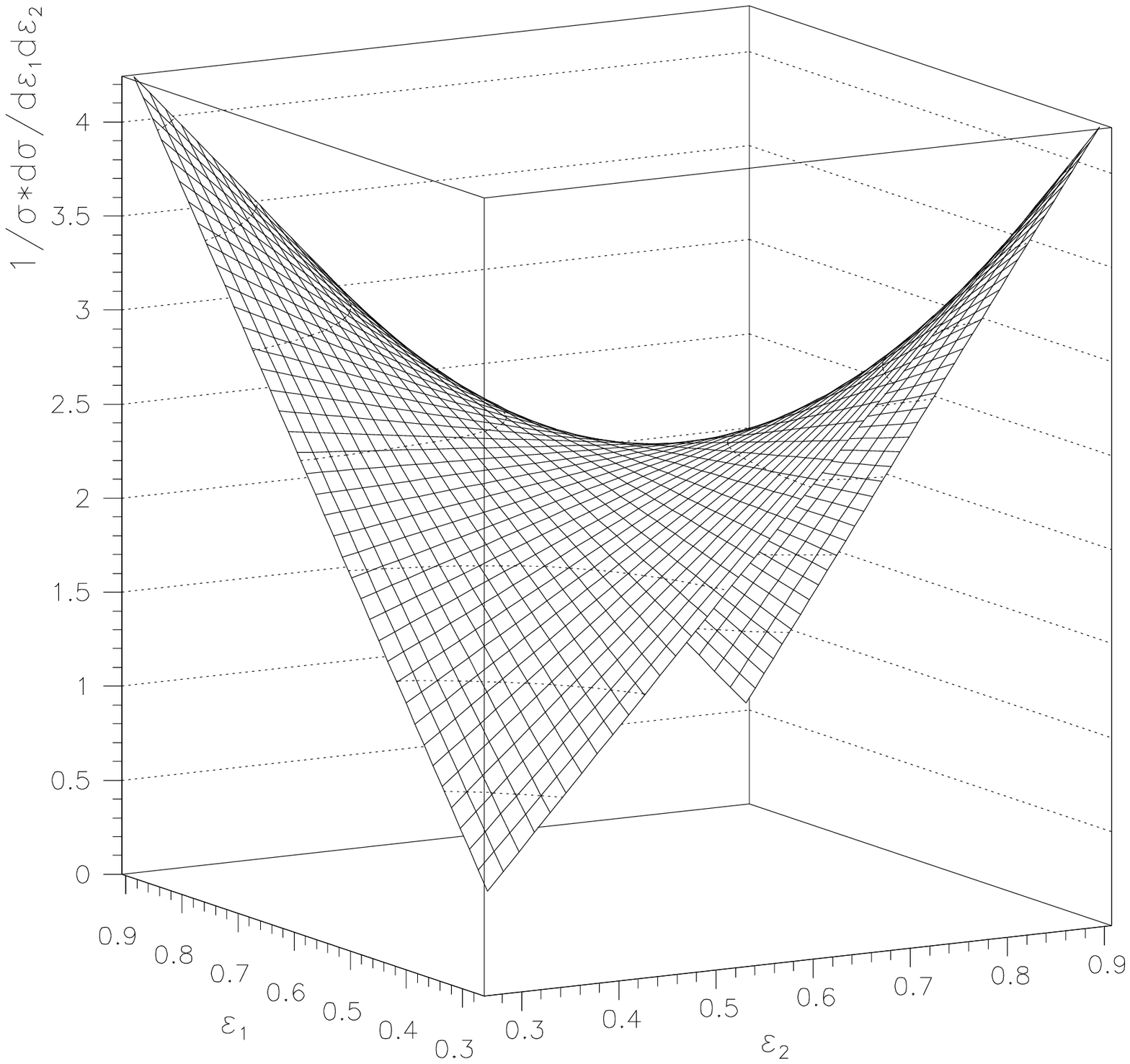}}
\mbox{\epsfysize 9.5cm \epsffile{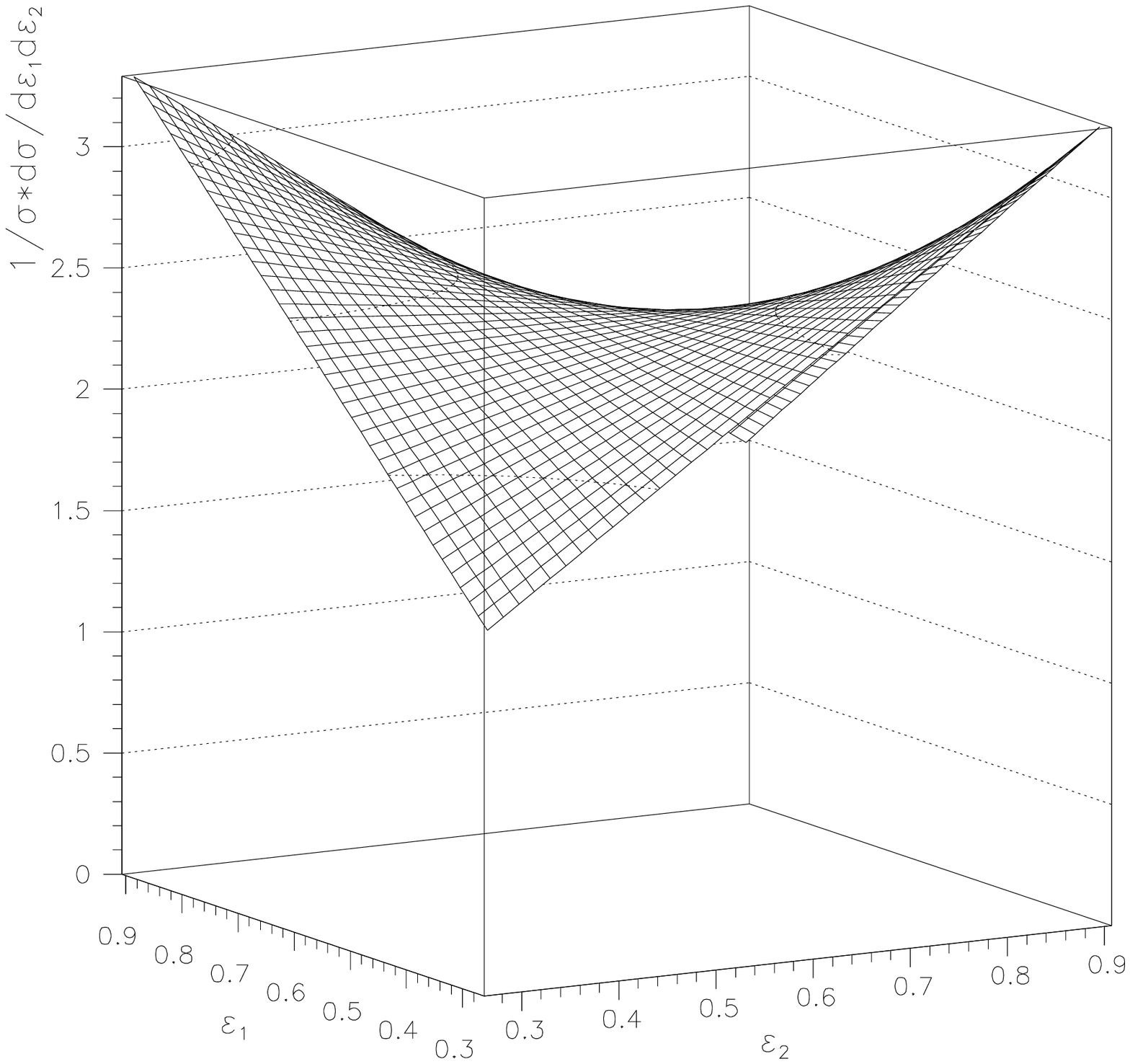}}
\caption{}
\end{figure}
\vspace{-15cm}
\qquad\qquad(a) \qquad\qquad\qquad\qquad
\qquad\qquad\qquad\qquad Majorana ($+$)
\newline

\vspace{7.5cm}
\qquad\qquad(b) \qquad\qquad\qquad\qquad
\qquad\qquad\qquad\qquad Majorana ($-$)}
\newpage
{\Large $\Emm$ Final State: Angular Correlation
\begin{figure}[H]
\mbox{\epsfysize 9.5cm \epsffile{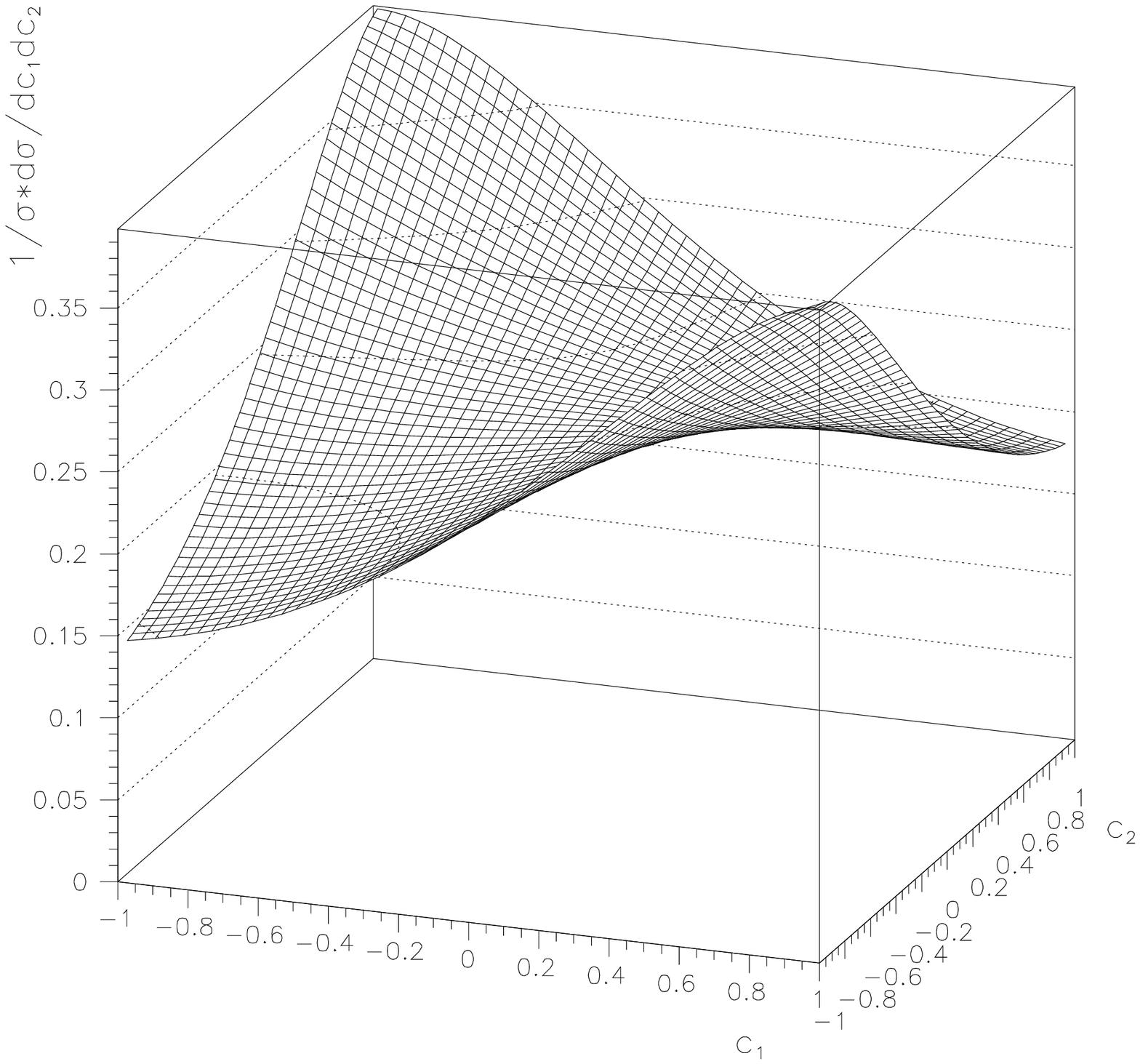}}
\mbox{\epsfysize 9.5cm \epsffile{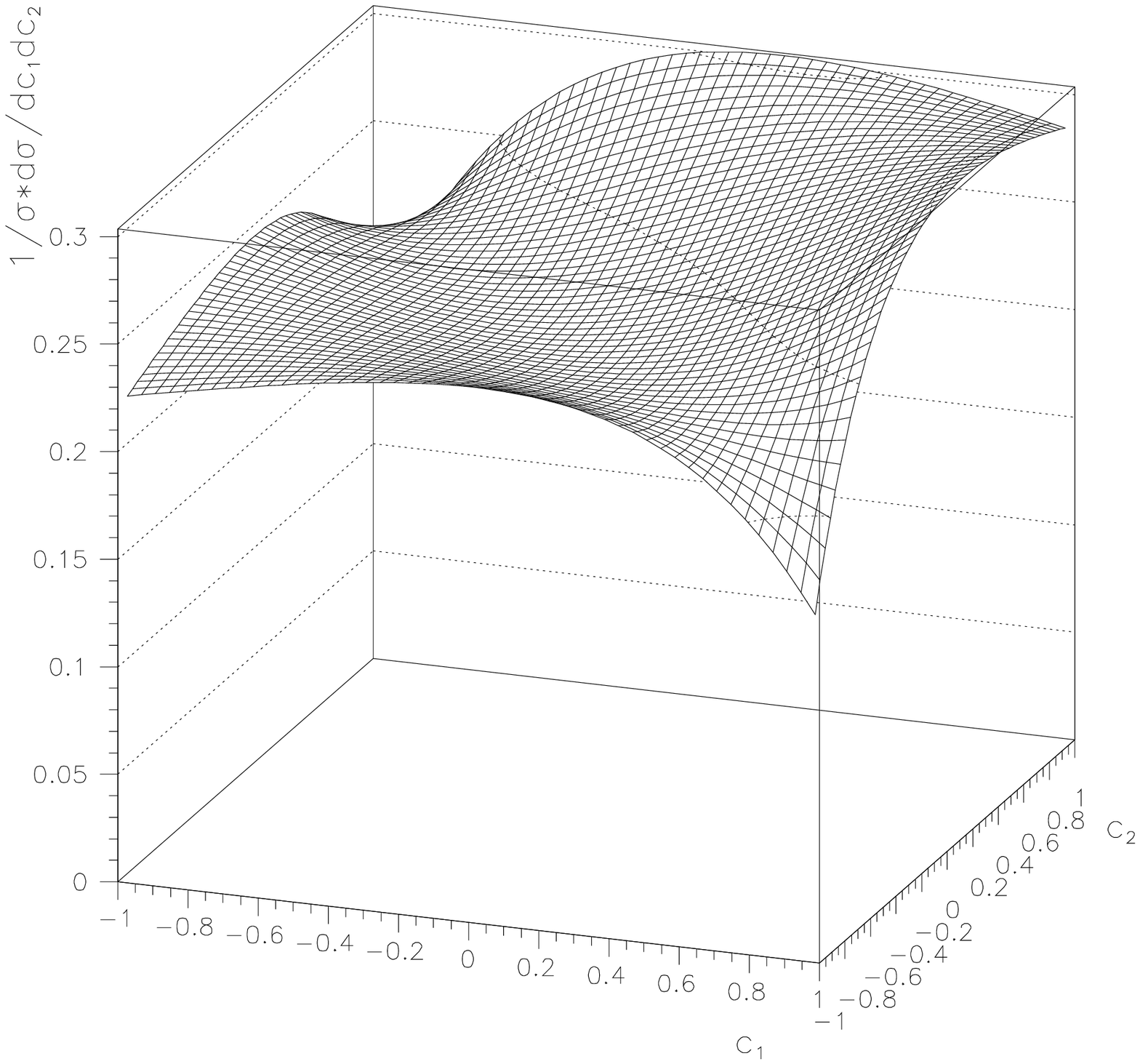}}
\caption{}
\end{figure}
\vspace{-15cm}
\qquad\qquad(a) \qquad\qquad\qquad\qquad
\qquad\qquad\qquad\qquad Majorana ($+$)
\newline

\vspace{7.5cm}
\qquad\qquad(b) \qquad\qquad\qquad\qquad
\qquad\qquad\qquad\qquad Majorana ($-$)}
\newpage
{\Large $\E$ Final State: Energy Correlation

\vspace{1cm}
Majorana ($+$)\qquad\qquad\qquad\qquad\qquad\qquad Majorana ($-$)
\begin{figure}[H]
\mbox{\epsfxsize 8cm \epsffile{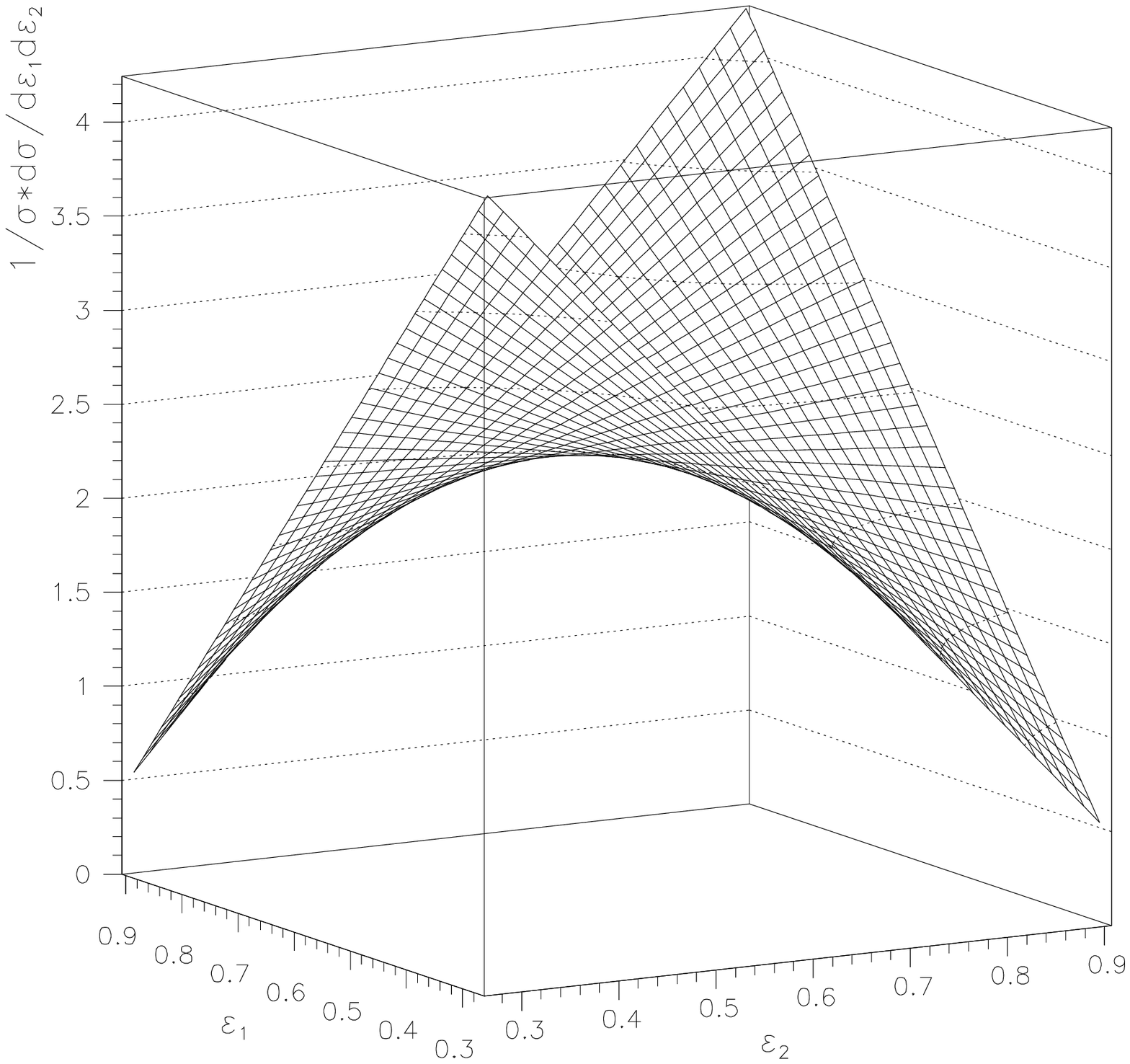}}
\mbox{\epsfysize 8cm \epsffile{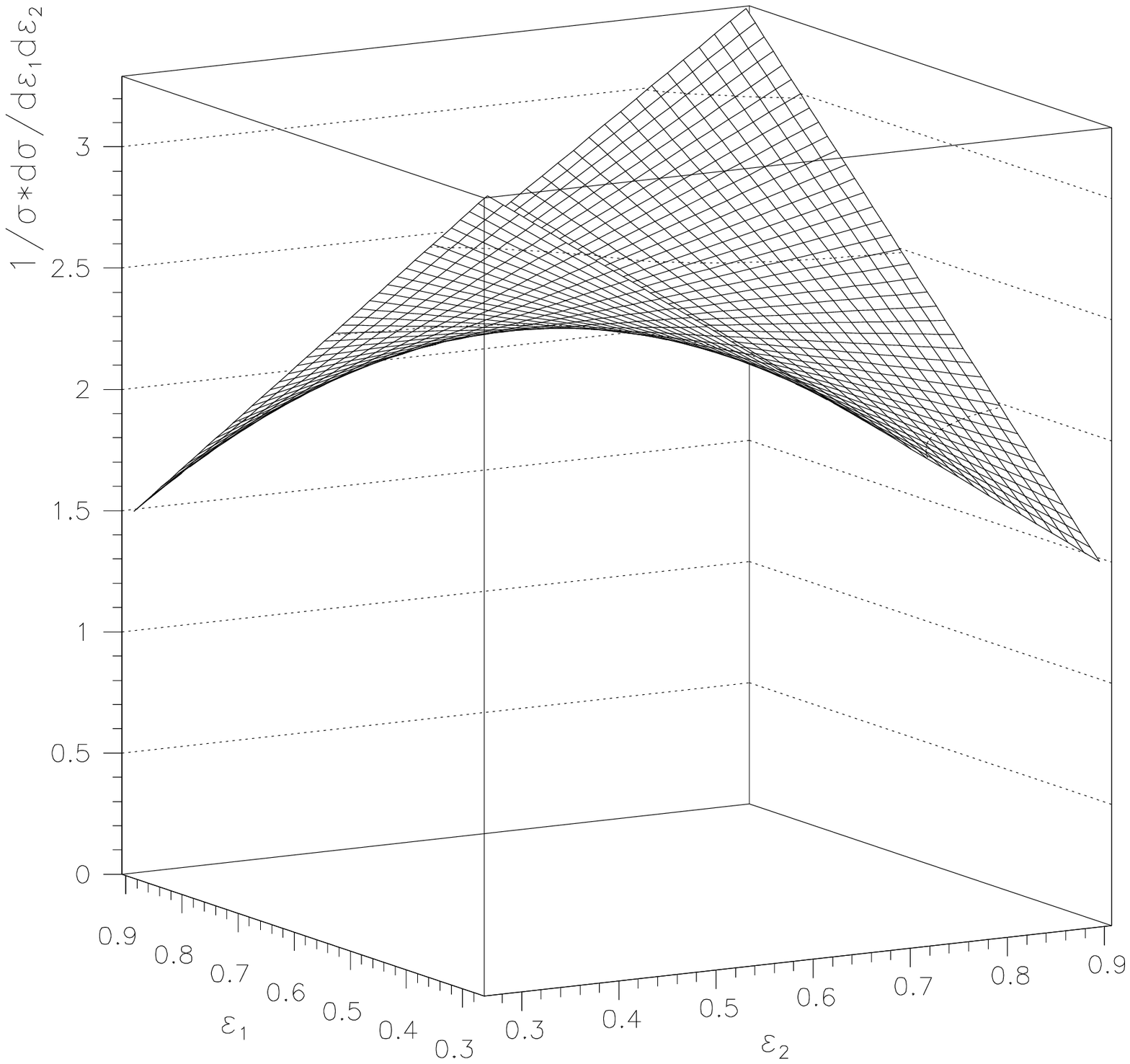}}
\end{figure}
 
\qquad Dirac}
\begin{figure}[b]
\mbox{\epsfysize 8cm \epsffile{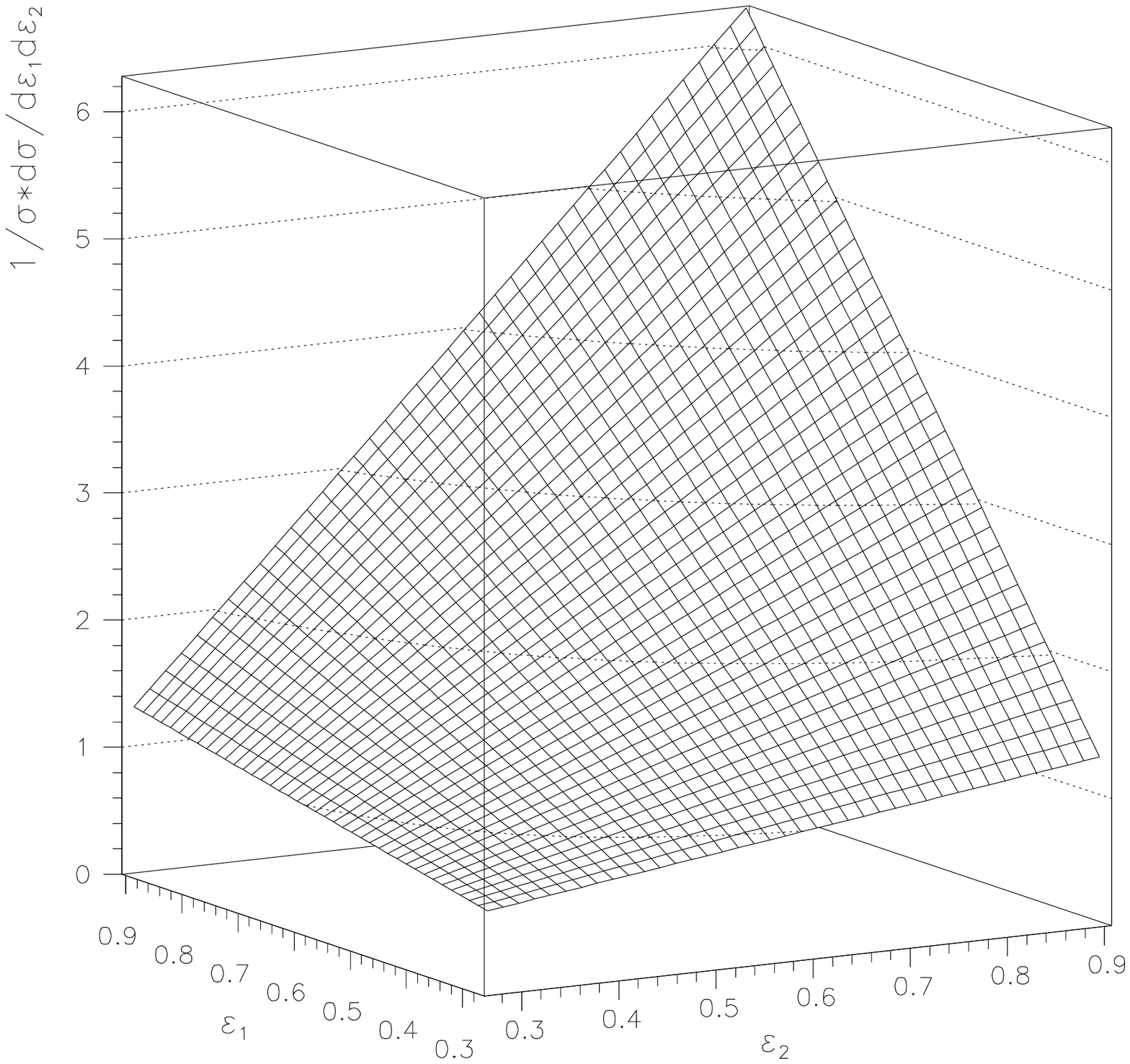}}
\caption{}
\end{figure}
\end{center}
\vspace{-12.5cm}
{\Large (a) \quad\qquad\qquad\qquad\qquad\qquad\qquad(b)
\newline

\vspace{7.5cm}
\qquad\qquad\qquad(c)}
\newpage
\begin{center}
{\Large $\E$ Final State: Angular Correlation

\vspace{1cm}
Majorana ($+$)\qquad\qquad\qquad\qquad\qquad\qquad Majorana ($-$)
\begin{figure}[H]
\mbox{\epsfysize 8cm \epsffile{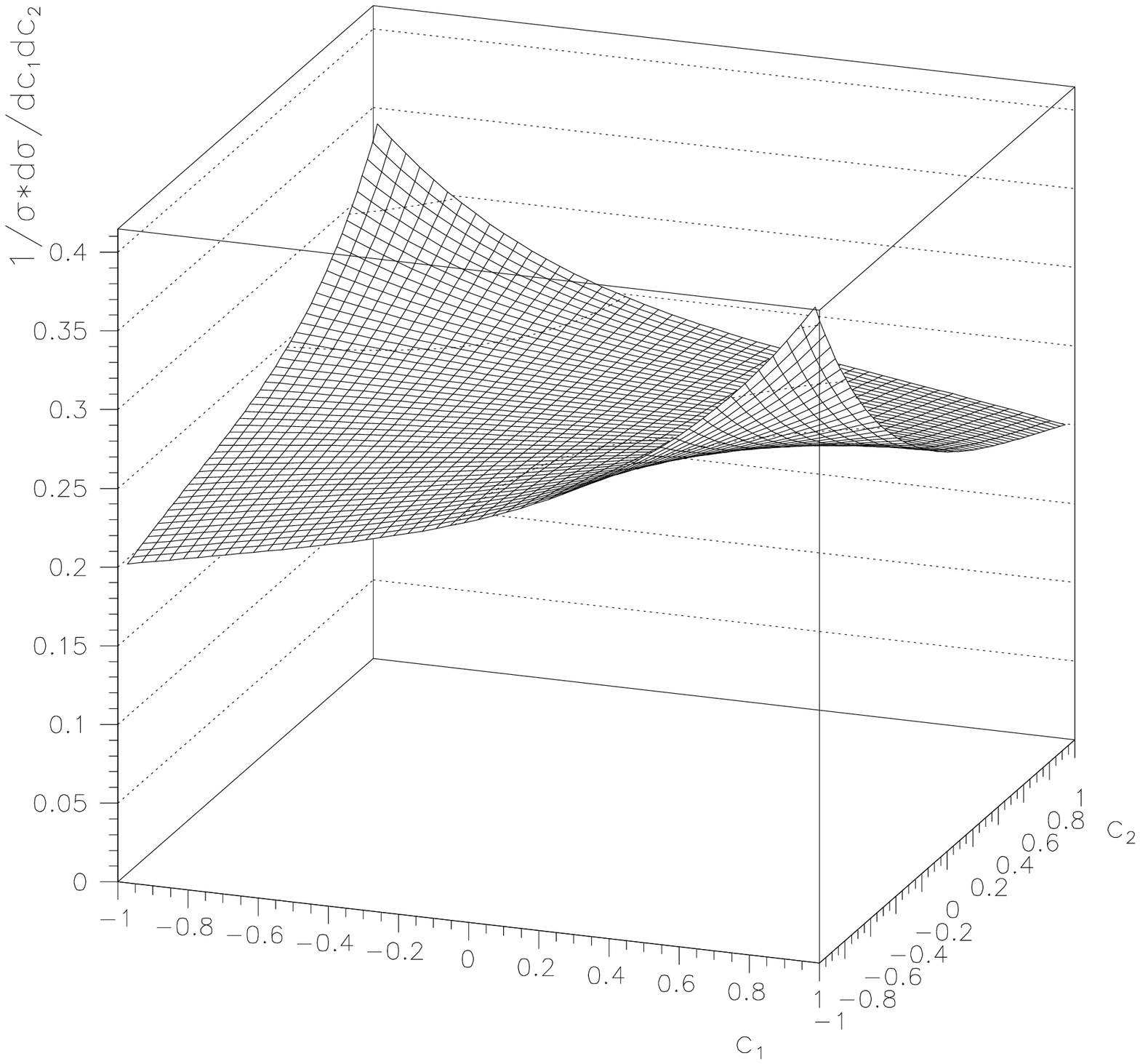}}
\mbox{\epsfysize 8cm \epsffile{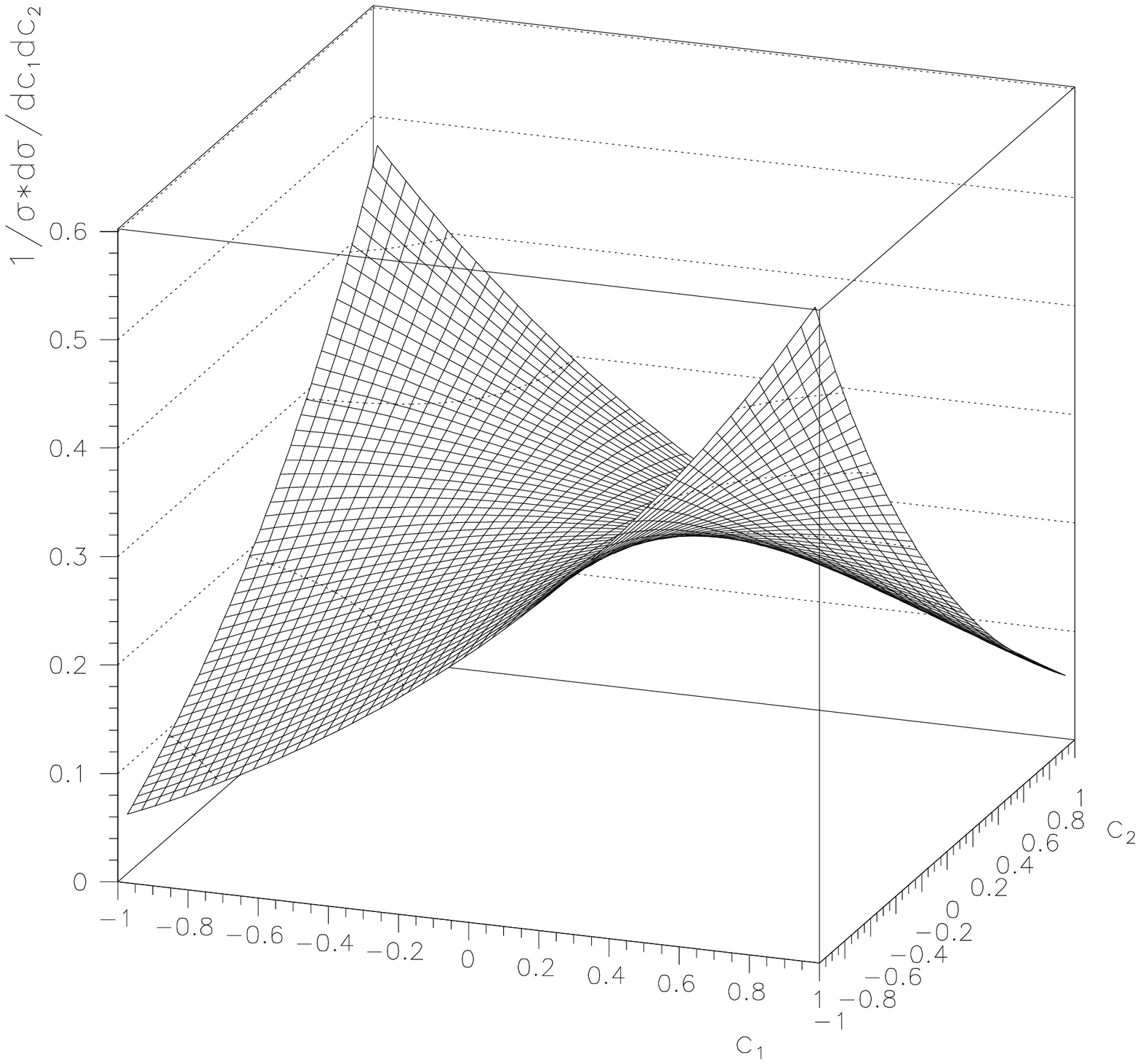}}
\end{figure} 

\qquad Dirac}
\begin{figure}[b]
\mbox{\epsfysize 8cm \epsffile{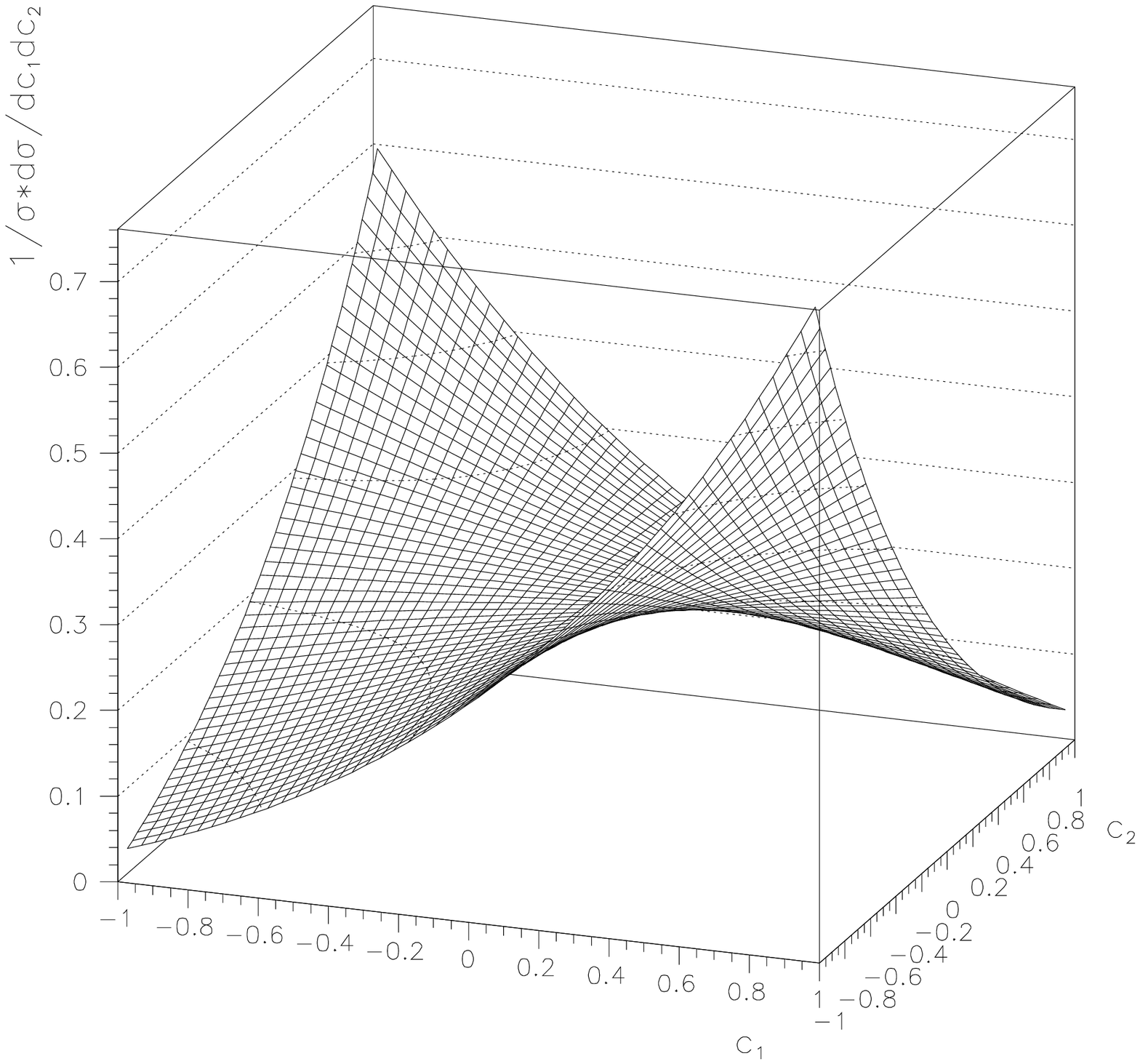}}
\caption{}
\end{figure}
\end{center} 
\vspace{-12.5cm}
{\Large (a) \qquad\qquad\qquad\qquad\qquad\qquad\qquad(b)
\newline

\vspace{7.5cm}
\qquad\qquad\qquad(c)}
\end{document}